\documentclass[10pt, conference, letterpaper]{IEEEtran}
\IEEEoverridecommandlockouts
\usepackage{cite}
\usepackage{amsmath,amssymb,amsfonts}
\usepackage{graphicx}
\usepackage{textcomp}
\usepackage{xcolor}
\usepackage{algorithm}
\usepackage{algpseudocode}
\usepackage{subcaption} 
\usepackage{booktabs}
\usepackage{url}
\def\BibTeX{{\rm B\kern-.05em{\sc i\kern-.025em b}\kern-.08em
    T\kern-.1667em\lower.7ex\hbox{E}\kern-.125emX}}

\title{SlicePilot: Demystifying Network Slice Placement in Heterogeneous Cloud Infrastructures}

\author{
\IEEEauthorblockN{Ioannis Panitsas\textsuperscript{*\textsection}, 
Tolga O.~Atalay\textsuperscript{\textsection}, 
Dragoslav Stojadinovic\textsuperscript{\textsection}, 
Angelos Stavrou\textsuperscript{\textdagger\textsection}, 
Leandros Tassiulas\textsuperscript{*}}
\IEEEauthorblockA{\textsuperscript{*}Department of Electrical and Computer Engineering, Yale University, New Haven, CT, USA}
\IEEEauthorblockA{\textsuperscript{\textdagger}Department of Electrical and Computer Engineering, Virginia Tech, Blacksburg, VA, USA}
\IEEEauthorblockA{\textsuperscript{\textsection}A2 Labs, LLC, Arlington, VA, USA}
}

\begin{document}
\maketitle

\begin{abstract} 
Cellular networks are comprised of software-based entities, with main functions encapsulated as Virtual Network Functions (VNFs) deployed on Commercial-off-the-Shelf (COTS) hardware. As a key enabler of 5G, network slicing offers logically isolated Quality of Service (QoS) for diverse use cases. With the transition to cloud-native infrastructures, optimizing network slice placement across multi-cloud environments remains challenging due to heterogeneous resource capabilities and varying slice-specific demands. This paper presents \textit{SlicePilot}, a modular framework that enables autonomous and near-optimal VNF placement using a disaggregated Multi-Agent Reinforcement Learning (MARL) approach. SlicePilot collects real-world traffic profiles to estimate resource needs for each slice type. These estimates guide a MARL-based scheduler that minimizes deployment costs while satisfying QoS constraints. We evaluate SlicePilot on a multi-cloud testbed and demonstrate a 19× speed-up over combinatorial optimization methods, while keeping deployment costs within 7.8\% of the optimal. Although SlicePilot results in 2.42× more QoS violations under high-load conditions, this trade-off is offset by faster decision-making and reduced computational overhead. Overall, SlicePilot delivers a scalable, cost-efficient solution for network slice placement, making it suitable for real-time deployments where responsiveness and efficiency are critical.

\end{abstract}

\section{Introduction}

Next-generation cellular networks are evolving 
to support various use cases, each with distinct Quality of Service (QoS) requirements. To accommodate this evolving diversity, ``network slicing''  has emerged as a key enabler, facilitating the creation of virtually isolated network segments over a shared physical infrastructure to support different use cases like URLLC (Ultra-Reliable Low-Latency Communication), eMBB (Enhanced Mobile Broadband), and mMTC (massive Machine Type Communications)~\cite{surv1,surv4}.

The advent of network slicing is facilitated by the adoption of foundational technologies such as Network Function Virtualization (NFV) and Software Defined Networking (SDN)~\cite{etsi_nfv_whitepaper, onf_sdn_whitepaper}. Leveraging NFV and SDN, the proprietary Physical Network Functions (PNFs) that made up the legacy 4G Radio Access Network (RAN) and Core Network (CN), are now deployed as Virtual Network Functions (VNFs) on Commercial-off-the-Shelf (COTS) hardware in the form of finer-grained microservices. A network slice is ultimately formed by service chaining a set of VNFs to create an isolated network segment that extends from the RAN to the CN \cite{surv4}. 

To capitalize on this software-driven approach, network operators are partnering with cloud providers to advance cloud-based 5G deployments~\cite{5GCore5G67online, 5GCoreon75online}. This shift enables the dynamic instantiation and migration of network slice VNFs across a geographically distributed cloud infrastructure. By leveraging this flexibility, operators can tailor deployment strategies to accommodate the diverse Service Level Agreements (SLAs) of different use cases. Specifically, operators can strategically position latency-sensitive VNFs closer to the RAN at edge locations~\cite{5GEdgeCo6online, LowLaten55online}, despite higher computational costs, while hosting less critical functions more cost-effectively in regional or central clouds~\cite{surv4, surv6, surv7}. Modern multi-cloud infrastructures provide a wide range of computing, networking, and storage capabilities. With infrastructure heterogeneity, coupled with diverse service requirements in terms of latency, bandwidth, and reliability, strategic VNF placement decisions become essential for optimizing resource use and supporting varied application demands in next-generation networks.

\textbf{Motivation}. The problem of network slice placement is widely recognized in the literature as a variant of the Virtual Network Embedding (VNE) problem, which is proven to be NP-hard ~\cite{surv6}. Prior work has explored Integer Linear Programming (ILP)-based methods to achieve optimality~\cite{opt1, opt2, opt3}, but these approaches do not scale and are computationally intensive, making them impractical for real-time decisions. Heuristic-based strategies have been proposed to improve efficiency, yet they often fail to achieve solutions close to the optimum, as they tend to get stuck in local optima ~\cite{opt5, opt4}. More recently, Reinforcement Learning (RL) has gained traction as a promising approach for optimal slice placement through trial and error procedures~\cite{rl4, rl6}. However, existing RL solutions predominantly rely on monolithic agent implementations that struggle to capture resource and service constraints of different slice types~\cite{rl1, rl2,rl3}, lack real-world validation~\cite{rl3, testbed2}, and scale poorly in large deployments~\cite{rl2,rl3}.

To address the optimal VNF placement problem, this paper introduces \textbf{SlicePilot}, a Multi-Agent Reinforcement Learning (MARL)-based placement framework designed to minimize human intervention and enable automated, zero-touch network management. \textit{SlicePilot} automates the deployment of network slice VNFs and disaggregated RAN components by leveraging slice-specific traffic profiles, resource demands, and infrastructure resource state. This information is then used to craft an optimal VNF placement strategy within a distributed cloud environment.  Building on this foundation and practical insights, this work makes the following key contributions:


\begin{itemize}  
    \item We collect real-world traffic from our over-the-air 5G testbed, capturing diverse slice-specific patterns to construct realistic and representative traffic profiles.

    \item We analyze the resource demands of key user plane VNFs by replaying the captured traffic, enabling accurate resource estimation and slice placement optimization through enhancements to the open-source OpenAirInterface (OAI) 5G software stack \cite{oai}.

    \item We design and implement a fully customized, disaggregated MARL scheduler that optimizes slice deployment in heterogeneous cloud infrastructures while jointly ensuring cost-efficiency and SLA compliance.

    \item We deploy \textit{SlicePilot} in a Kubernetes-orchestrated cluster hosted on OpenStack, emulating a realistic multi-cloud infrastructure environment. Our evaluation demonstrates that \textit{SlicePilot} delivers a 19× speed-up over traditional combinatorial optimization methods, achieves deployment costs within 7.8\% of the optimal solution, and tolerates 2.42× more SLA violations, highlighting its ability to enable cost-efficient and scalable slice deployments.

\end{itemize}

\noindent The remainder of the paper is organized as follows. Section II provides background on 5G architecture and network slicing. Section III introduces the \textit{SlicePilot} framework along with its components. Section IV outlines the experimental setup used for real-world validation while Section V presents the evaluation results, followed by a discussion of related work in Section VI. Finally, Section VII concludes our work.

\section{BACKGROUND}
\begin{figure}[t]
    \centering
    \includegraphics[width=0.37\textwidth]{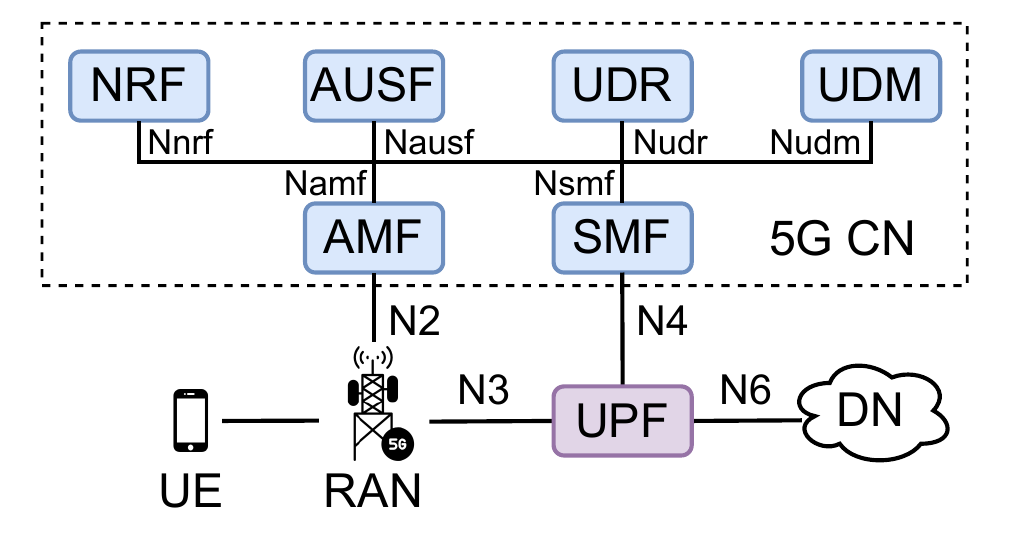}
    \caption{5G system architecture.}
    \label{fig:5g_system_architecture}
    \vspace{-10pt} 
\end{figure}

\noindent \textbf{5G System Architecture}. A 5G cellular system, as illustrated in Fig.~\ref{fig:5g_system_architecture}, consists of three main components: the RAN, which provides over-the-air connectivity between User Equipment (UE) and the network; the CN, which manages session control, mobility, and interconnectivity with external Data Networks (DNs); and the Transport Network (TN), responsible for carrying traffic between RAN and CN elements.

\noindent \textbf{Radio Access Network}:  Unlike previous generations, the 5G RAN embraces a disaggregated virtualized architecture that enhances efficiency and scalability by allowing network operators to dynamically allocate resources based on service demands~\cite{polese2023understanding}. This is achieved by virtualizing baseband processing functions, enabling flexible deployment on cloud infrastructures. The RAN is composed of three main units: the Radio Unit (RU), which handles radio transmission, reception, and antenna processing; the Distributed Unit (DU), responsible for lower-layer processing such as error correction and resource allocation; and the Centralized Unit (CU), which manages higher-layer functions~\cite{polese2023understanding}.

\noindent \textbf{Core Network}: The 5G CN is designed with a service-based, cloud-native architecture to ensure modularity, scalability, and support for network slicing \cite{3GPP_TS_23_501}. It is composed of several control plane VNFs that operate on a consumer-producer model, communicating with each other via standardized REST APIs. At the heart of the 5G Core is the Access and Mobility Management Function (AMF), which acts as the communication hub between the RAN, UE, and other VNFs, providing essential signaling and mobility management services. The Authentication Server Function (AUSF) provides authentication credentials to the AMF, working in conjunction with the Unified Data Management (UDM) and Unified Data Repository (UDR) to implement the 5G Authentication and Key Agreement (AKA) procedure. Within the CN, the Session Management Function (SMF) and User Plane Function (UPF) serve as the control and user plane anchors, respectively. The UPF tunnels user traffic toward an external DN node, which represents real-world application servers. Finally, the Network Repository Function (NRF) facilitates initial discovery and communication among VNFs.

\noindent \textbf{Network Slicing}. A network slice is an end-to-end logical network instance formed by service chaining a set of VNFs, including both control and user plane components, to meet specific application requirements. Each VNF has distinct resource requirements, and is deployed across a virtualized infrastructure. While network slicing enables unprecedented flexibility by allowing operators to create isolated, end-to-end virtual networks, it also introduces complex challenges in orchestration, resource allocation, and slice placement optimization \cite{surv4}. Several standardization efforts have been proposed to define and manage the lifecycle of network slices. Among them, the ETSI NFV architecture provides a comprehensive framework for deploying VNFs, and automating slice operations~\cite{etsi_nfveve012v311}. At its core, the Network Function Virtualization Orchestrator (NFVO) coordinates slice instantiation, scaling, and lifecycle management. The Virtualized Infrastructure Manager (VIM) oversees computing, storage, and networking resources, translating NFVO decisions into infrastructure-level actions. Beneath these layers, the Network Functions Virtualization Infrastructure (NFVI) unifies physical and virtual resources across edge, distributed, and central cloud domains, providing the foundation for hosting network slices.

\section{SLICEPILOT FRAMEWORK}

\textit{SlicePilot} is a modular framework designed to optimize network slice placement through a multi-stage approach. It consists of three functional blocks: (1) the User Plane Traffic Capturing Block (UPTCB), which collects and profiles over-the-air user plane traffic patterns for various slice types; (2) the User Plane VNF Stressing Block (UPVSB), which estimates VNF resource demands across different slice types; and (3) the Slice Placement Block (SPB), which optimizes slice deployment by leveraging slice-specialized RL agents to minimize network operator costs while ensuring SLA compliance.

\subsection{User Plane Traffic Capturing Block}

\begin{figure}[t]
    \centering
    \includegraphics[scale=0.41] {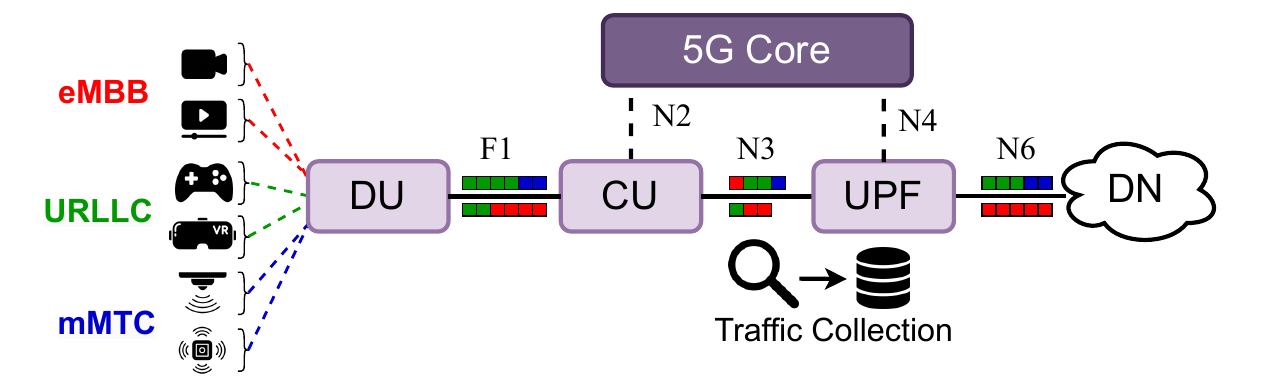} 
    \vspace{0.1pt} 
    \caption{UPTCB: Collecting and profiling over-the-air slice-specific user plane traffic for realistic replay and analysis.}
    \label{fig:user_plane_capturing}
\end{figure}

The UPTCB is the first building block of the \textit{SlicePilot} framework, designed to collect, preprocess, and profile over-the-air user plane traffic, enabling realistic slice performance analysis and optimization. Unlike synthetic traffic generators such as iPerf3 \cite{iperf3}, which offer controlled traffic generation but lack configurability for complex application-specific patterns, UPTCB builds a structured repository of preprocessed flows aligned with the three primary 5G slice types: eMBB for bandwidth-intensive applications, URLLC for ultra-low latency services, and mMTC for massive IoT deployments. By replaying these traffic profiles through the user plane path, as illustrated in Figure~\ref{fig:user_plane_capturing}, UPTCB enables accurate VNF resource estimation under real-world conditions, supporting informed slice placement and scaling. For interested readers, details on the traffic collection methodology are provided in Section \S\ref{sec:experimentation}.

\begin{figure*}[ht]
    \centering
    \noindent
    \begin{minipage}{0.16\textwidth} 
        \centering
        \includegraphics[width=1\textwidth]{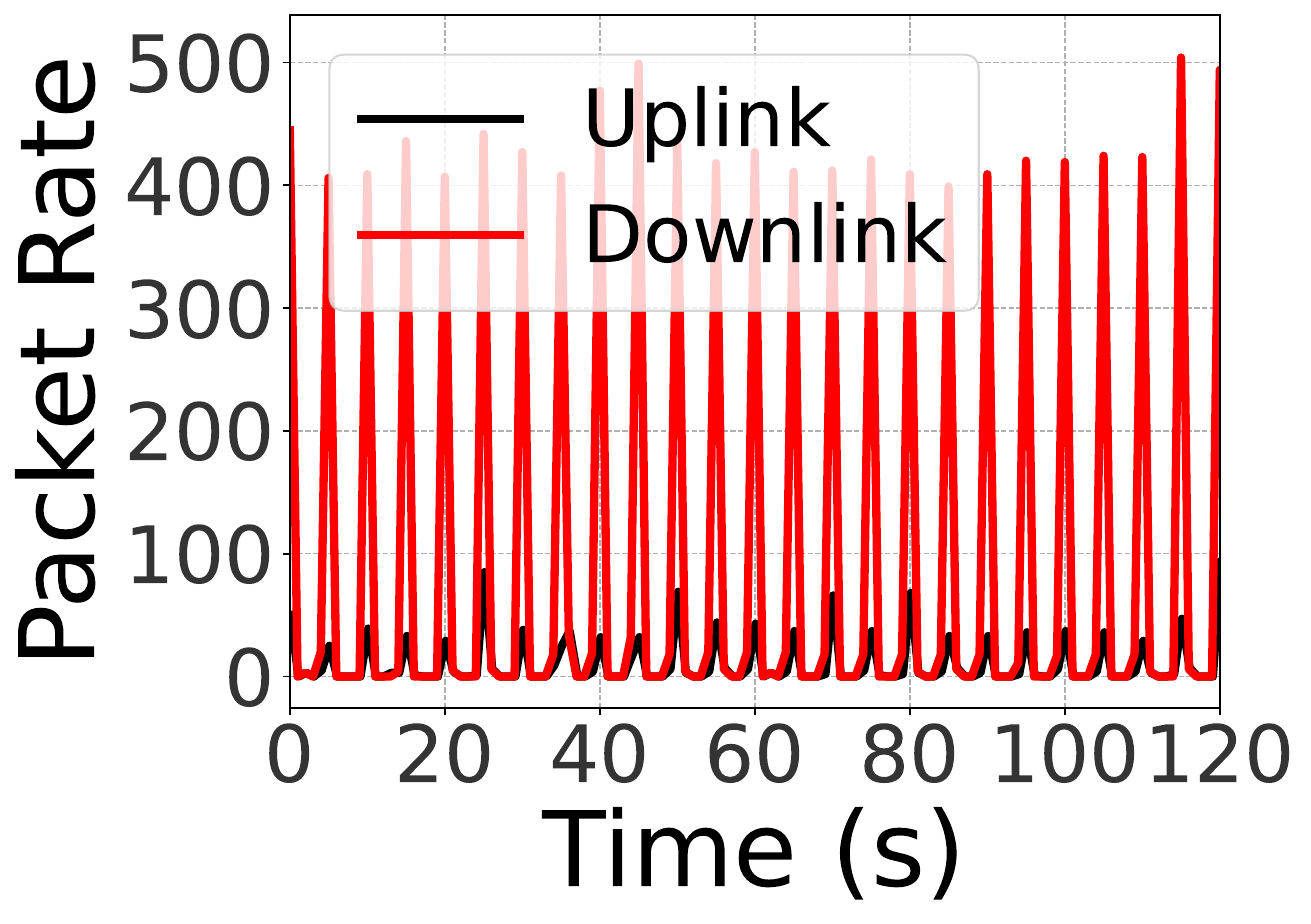} 
    \end{minipage}
    \begin{minipage}{0.16\textwidth}
        \centering
        \includegraphics[width=1\textwidth]{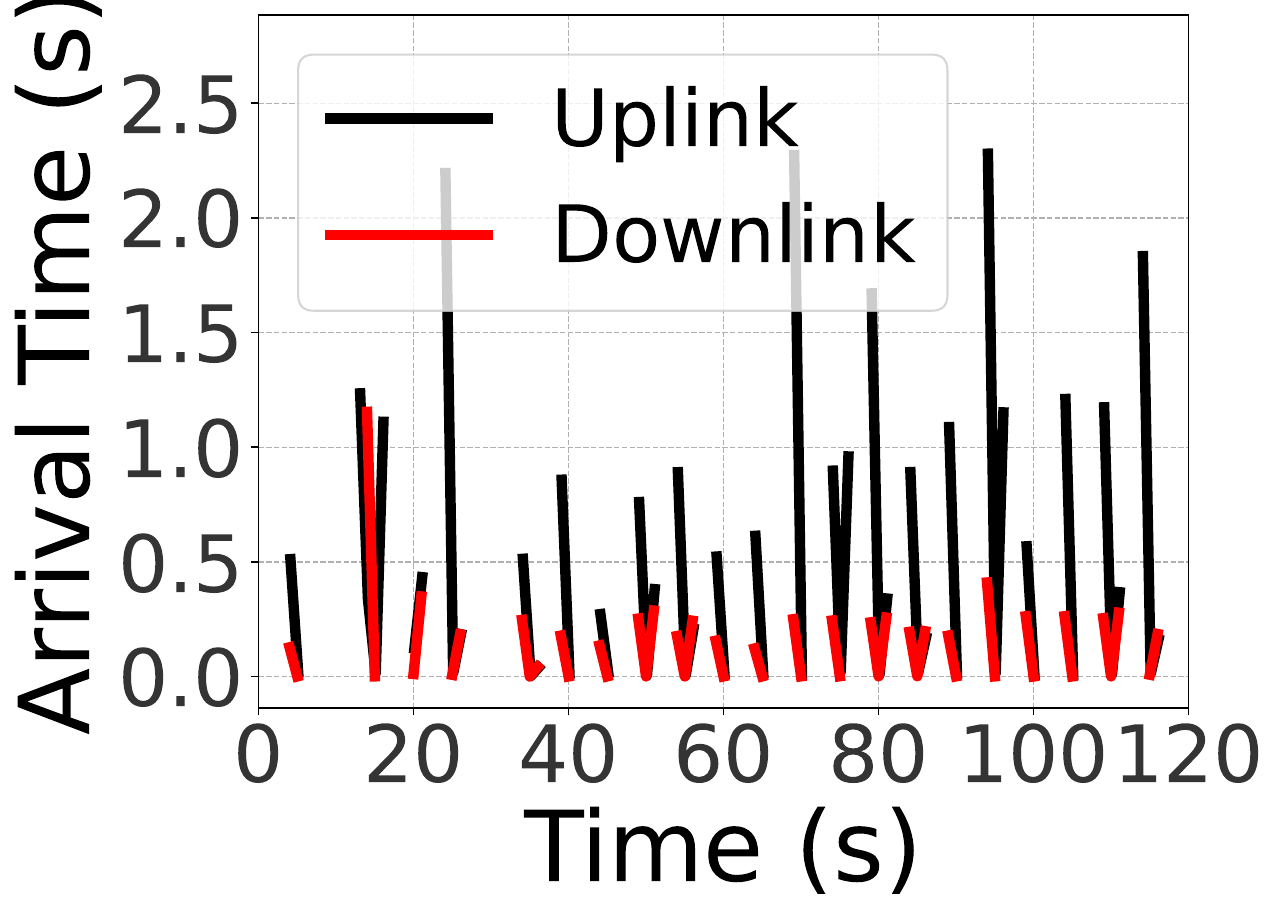}
    \end{minipage}
    \begin{minipage}{0.16\textwidth} 
        \centering
        \includegraphics[width=1\textwidth]{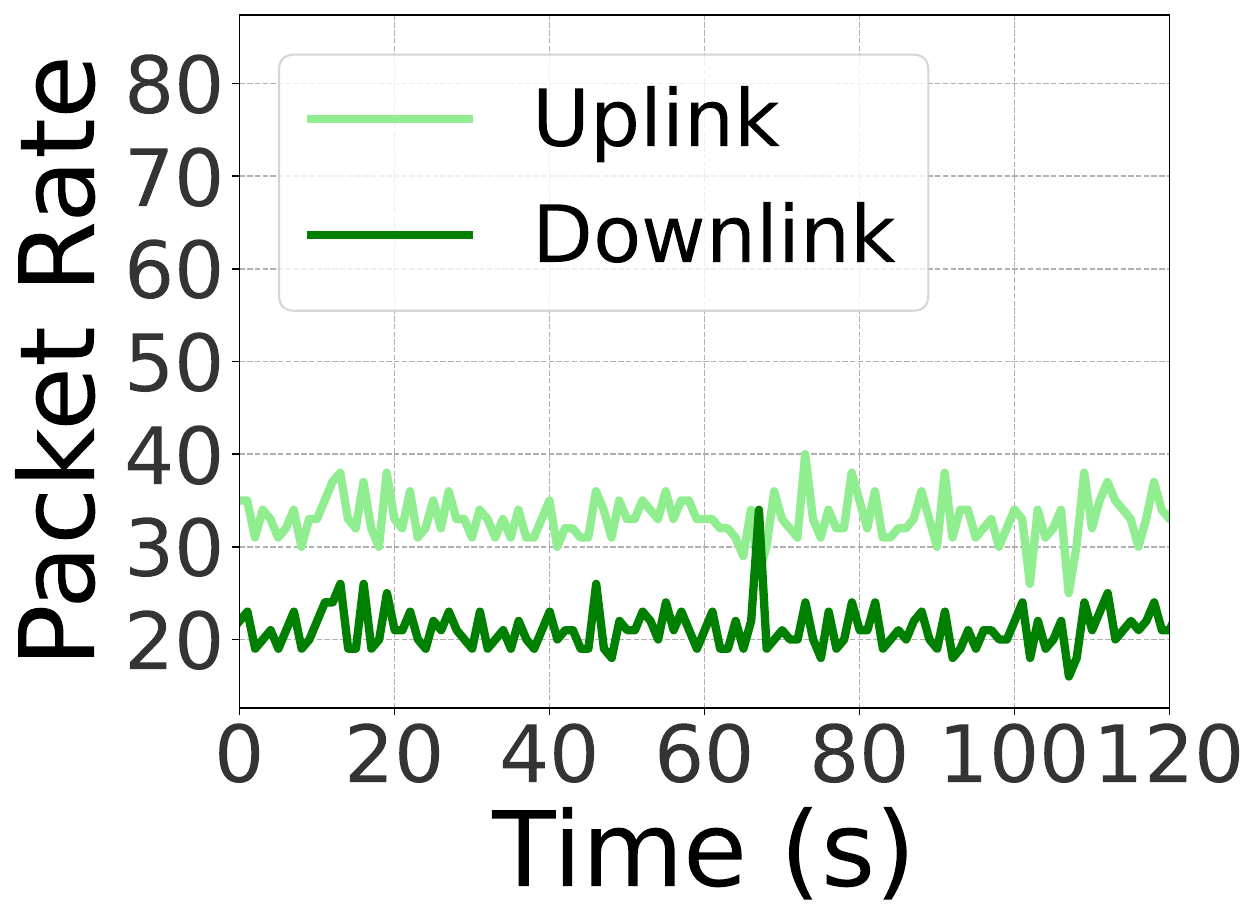}
    \end{minipage}
    \begin{minipage}{0.16\textwidth}
        \centering
        \includegraphics[width=1\textwidth]{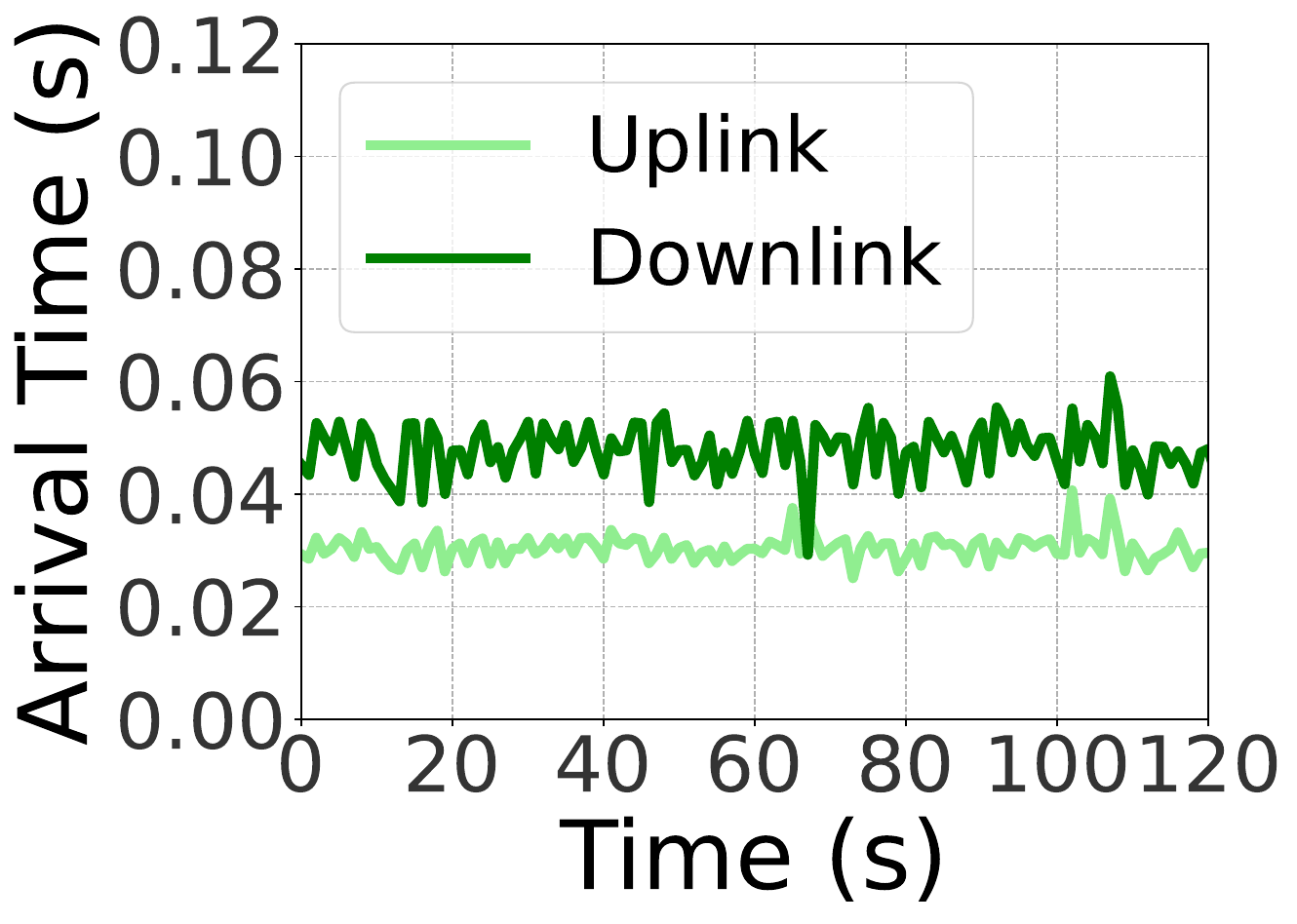}
    \end{minipage}
    \begin{minipage}{0.16\textwidth} 
        \centering
        \includegraphics[width=1\textwidth]{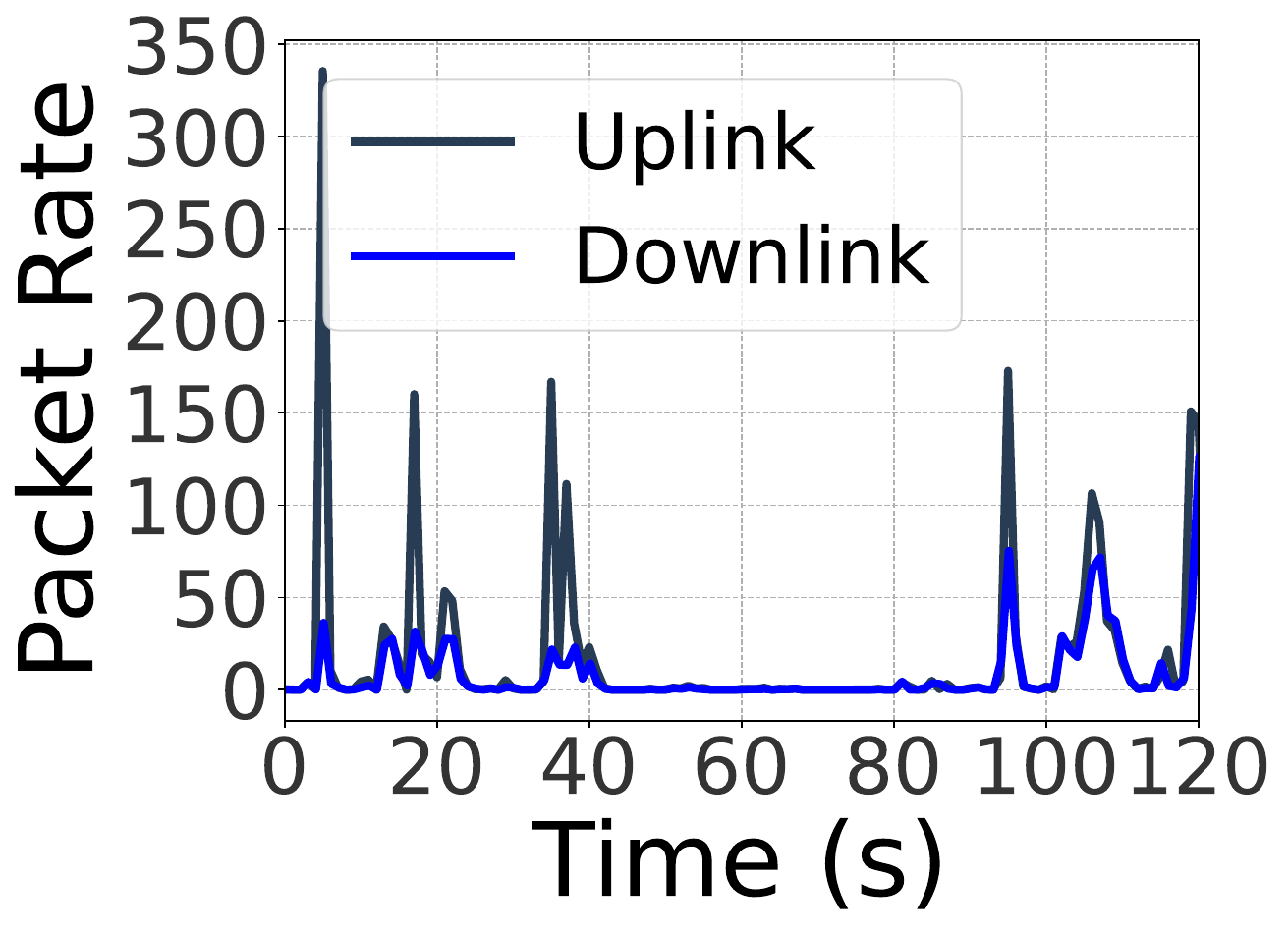}
    \end{minipage}
    \begin{minipage}{0.16\textwidth}
        \centering
        \includegraphics[width=1\textwidth]{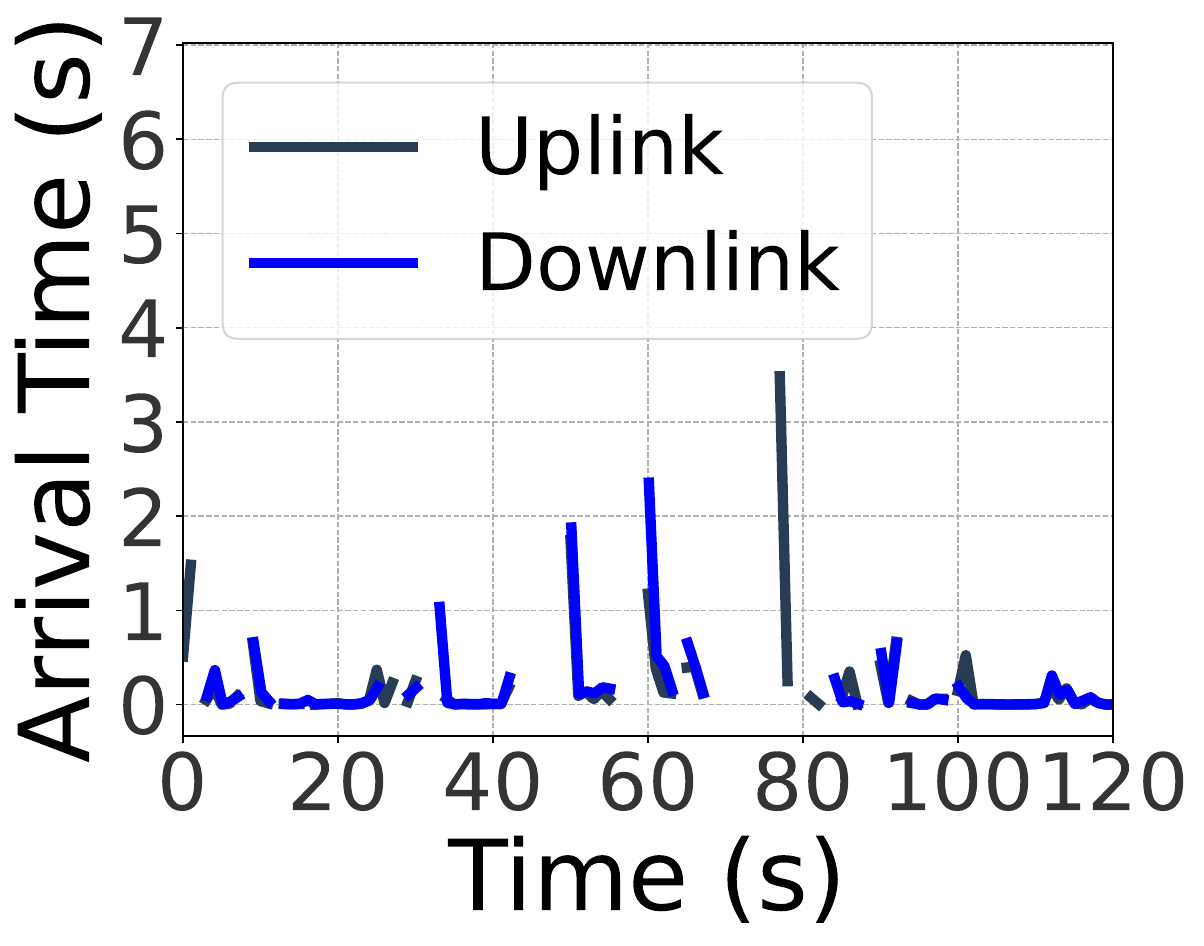}
    \end{minipage}

    \vspace{0.01cm} 
    \begin{minipage}{0.32\textwidth}
        \centering
        \text{\small eMBB Traffic (video streaming)} 
    \end{minipage}
    \begin{minipage}{0.32\textwidth}
        \centering
        \text{\small URLLC Traffic (online gaming)} 
    \end{minipage}
    \begin{minipage}{0.32\textwidth}
        \centering
        \text{\small mMTC Traffic (sensor connectivity)} 
    \end{minipage}


    \caption{Traffic profiling of eMBB, URLLC, and mMTC slices, illustrating packet rate (left) and inter-arrival time (right).}
    \label{fig:traffic_characteristics}

\end{figure*}

To illustrate the diversity of traffic patterns captured by UPTCB, Figure~\ref{fig:traffic_characteristics} presents three representative traces—one for each 5G slice type—highlighting both packet rate and inter-arrival time. The first pair of plots illustrates traffic from an eMBB slice (video streaming via YouTube), which exhibits periodic high-throughput bursts and longer inter-packet intervals, with a strong downlink dominance. The second pair corresponds to URLLC traffic (online gaming via Fortnite), where smaller payloads and shorter inter-arrival times reflect the low-latency, interactive nature of the application. The final pair represents mMTC traffic from a connected IoT sensor, characterized by infrequent transmissions, small payload sizes, and low sustained load. These distinct traffic patterns underscore the importance of slice-specific traffic profiling, as each slice type exhibits unique characteristics that demand different computational and network resources. Accurate profiling enables precise resource estimation and supports informed, adaptive, and efficient placement and scaling decisions.

\subsection{User Plane VNF Stressing Block}

\begin{figure}[t]
    \centering
    \includegraphics[scale=0.3]{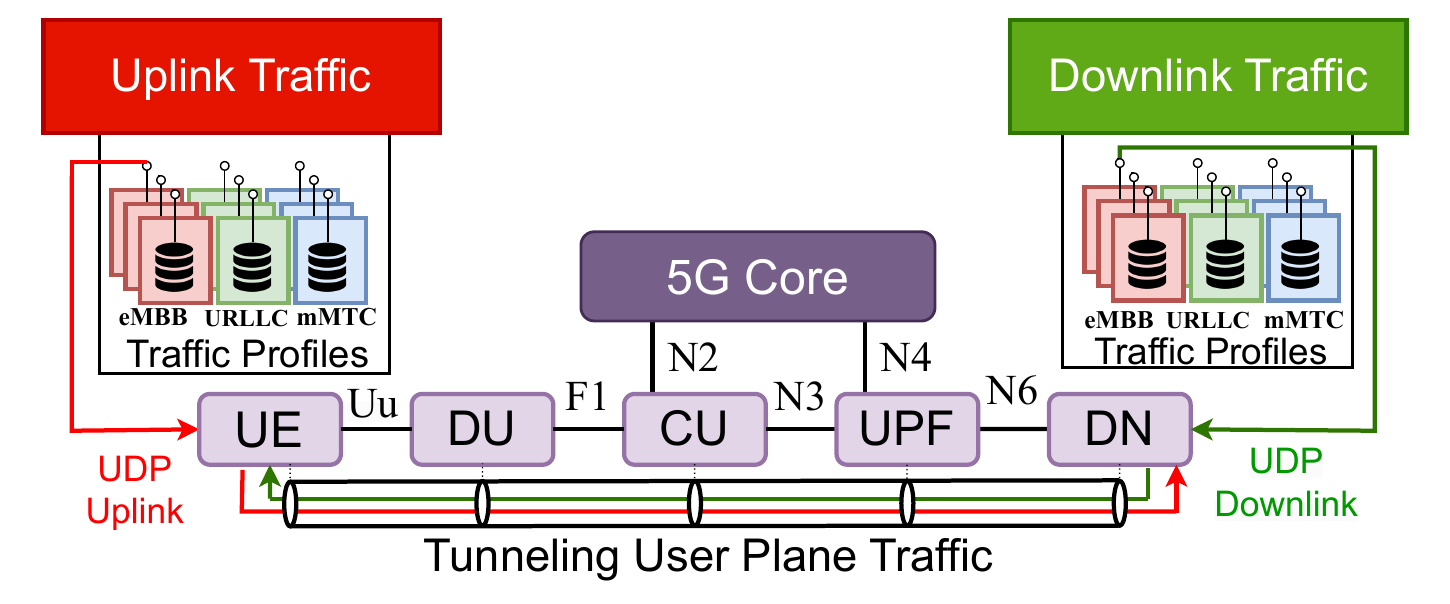}
    \vspace{2pt} 
    \caption{UPVSB: Replaying slice-specific traffic through the user plane for VNF resource profiling.}
    \label{fig:vnf_stressing}
    \vspace{-10pt}
\end{figure}

\begin{figure*}[t]
    \centering
    \noindent
    \begin{minipage}{0.32\textwidth}
        \centering
        \includegraphics[height=2.4cm]{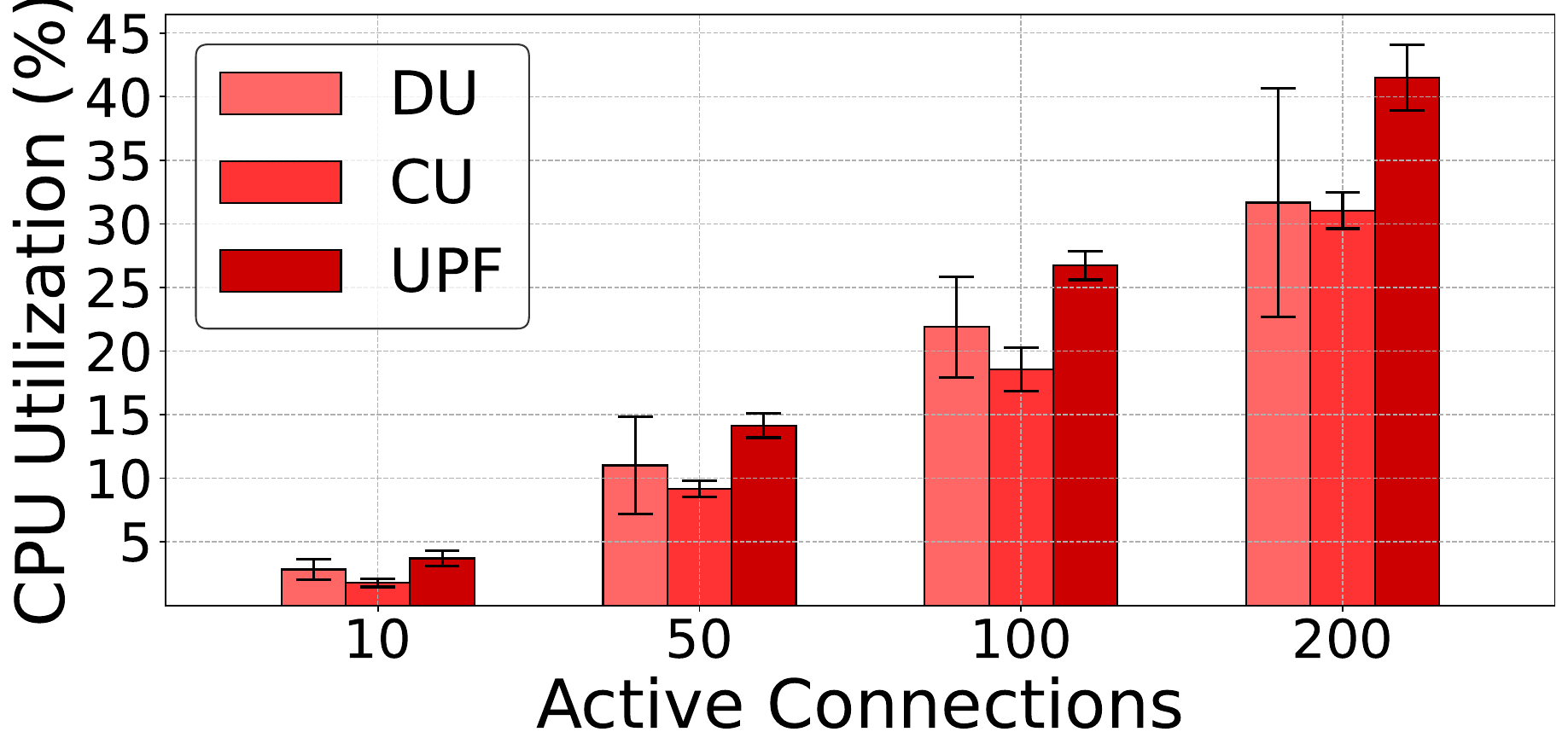}
    \end{minipage}
    \begin{minipage}{0.32\textwidth}
        \centering
        \includegraphics[height=2.4cm]{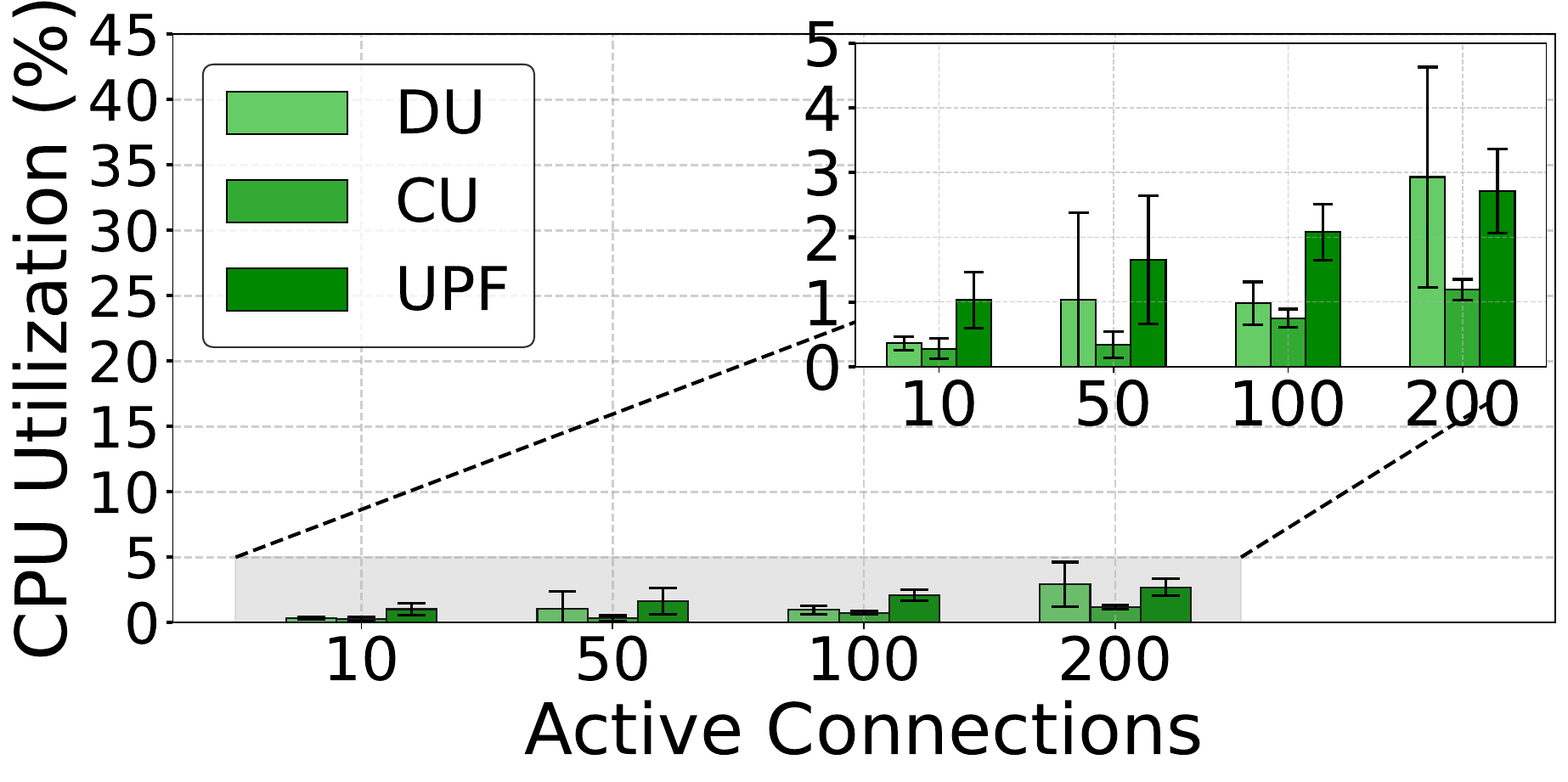}
    \end{minipage}
    \begin{minipage}{0.32\textwidth}
        \centering
        \includegraphics[height=2.4cm]{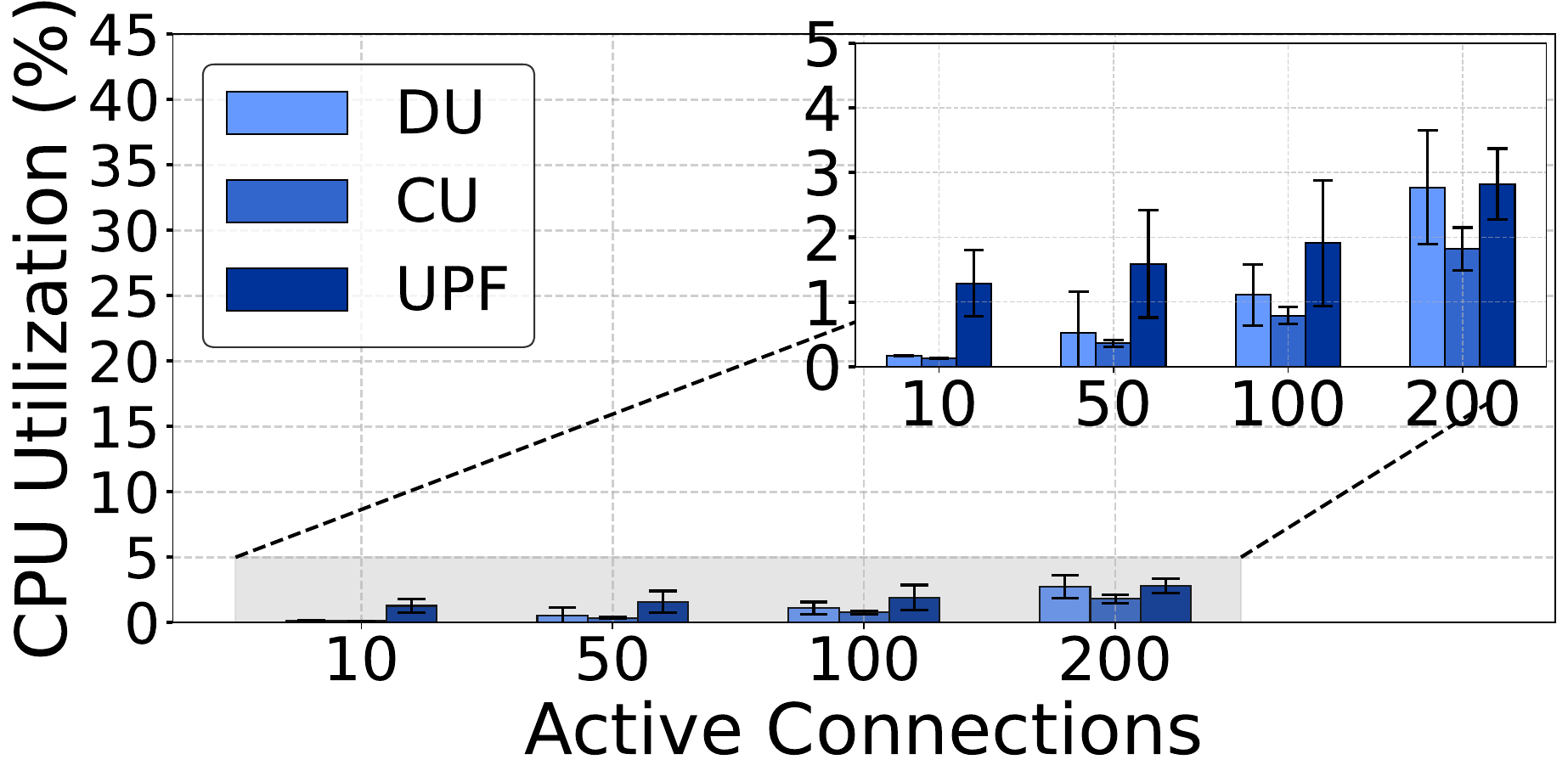}
    \end{minipage}
    \caption{User plane VNF resource usage for eMBB (red), URLLC (green), and mMTC (blue) slices under increasing user load.}
   \label{fig:vnf_stressing_measurements}
\end{figure*}

The UPVSB is the second component of the \textit{SlicePilot} framework, responsible for accurately estimating the resource requirements of each user plane VNF under different slice configurations and operational conditions. By applying controlled stress across varying load levels and slice types, UPVSB evaluates resource consumption and performance impact, providing critical insights for optimal VNF placement, capacity provisioning, and efficient network orchestration. As illustrated in Figure~\ref{fig:vnf_stressing}, UPVSB replays traffic profiles collected by UPTCB through the user plane path to systematically quantify per-slice resource demands, measuring the resulting computational and storage footprint across VNFs. This process produces a lookup table that captures upper bounds on resource usage as a function of slice type, service type, and the number of active user sessions, supporting rapid runtime decisions. UPVSB is implemented on top of the OAI 5G software stack, and further implementation details are provided in Section~\S\ref{sec:experimentation}.

To showcase UPVSB’s resource estimation methodology, we examine CPU utilization variations across slice types as the number of active user connections steadily increases over time. As shown in Figure~\ref{fig:vnf_stressing_measurements}, we evaluate resource utilization across user plane VNFs, including the DU, CU, and UPF, for three different slices—one from each category: eMBB, URLLC, and mMTC. CPU utilization scales sharply with user load for the eMBB slice, which supports a video streaming application. With 200 active connections, DU utilization rises to 31.67\%, CU utilization to 31.04\%, and UPF utilization peaks at 41.48\%, reflecting the processing overhead induced by high-throughput data plane traffic. Even at a lower load of 50 active connections, DU, CU, and UPF utilization already reach 11.01\%, 9.18\%, and 14.15\%, respectively, emphasizing the significant resource demands of high-bandwidth services. In contrast, the URLLC slice, customized for an online gaming application (Fortnite), maintains significantly lower CPU utilization. With 200 active connections, DU utilization remains at just 2.93\%, while CU and UPF utilization reach only 1.19\% and 2.71\%, respectively. This reflects the small packet sizes and stringent latency constraints of real-time applications, which require rapid but lightweight processing rather than heavy data forwarding. The mMTC slice, tailored for IoT sensor connectivity, exhibits a more gradual increase in CPU utilization as user connections grow. When the number of connected devices reaches 200, DU utilization is 2.77\%, CU utilization reaches 1.82\%, and UPF utilization remains at 2.82\%, demonstrating a steady yet lower resource footprint compared to eMBB. Even at the lowest tested load of 10 connections, all user plane VNFs maintain utilization below 1.5\%, reinforcing the low computational demand of IoT-based communication. Based on these insights, the next step is determining how to strategically place VNFs to efficiently accommodate diverse traffic demands while ensuring optimal network performance, resource utilization, and cost-effectiveness.

\subsection{Slice Placement Block}
\begin{figure}[t]
    \centering
    \includegraphics[scale=0.4]{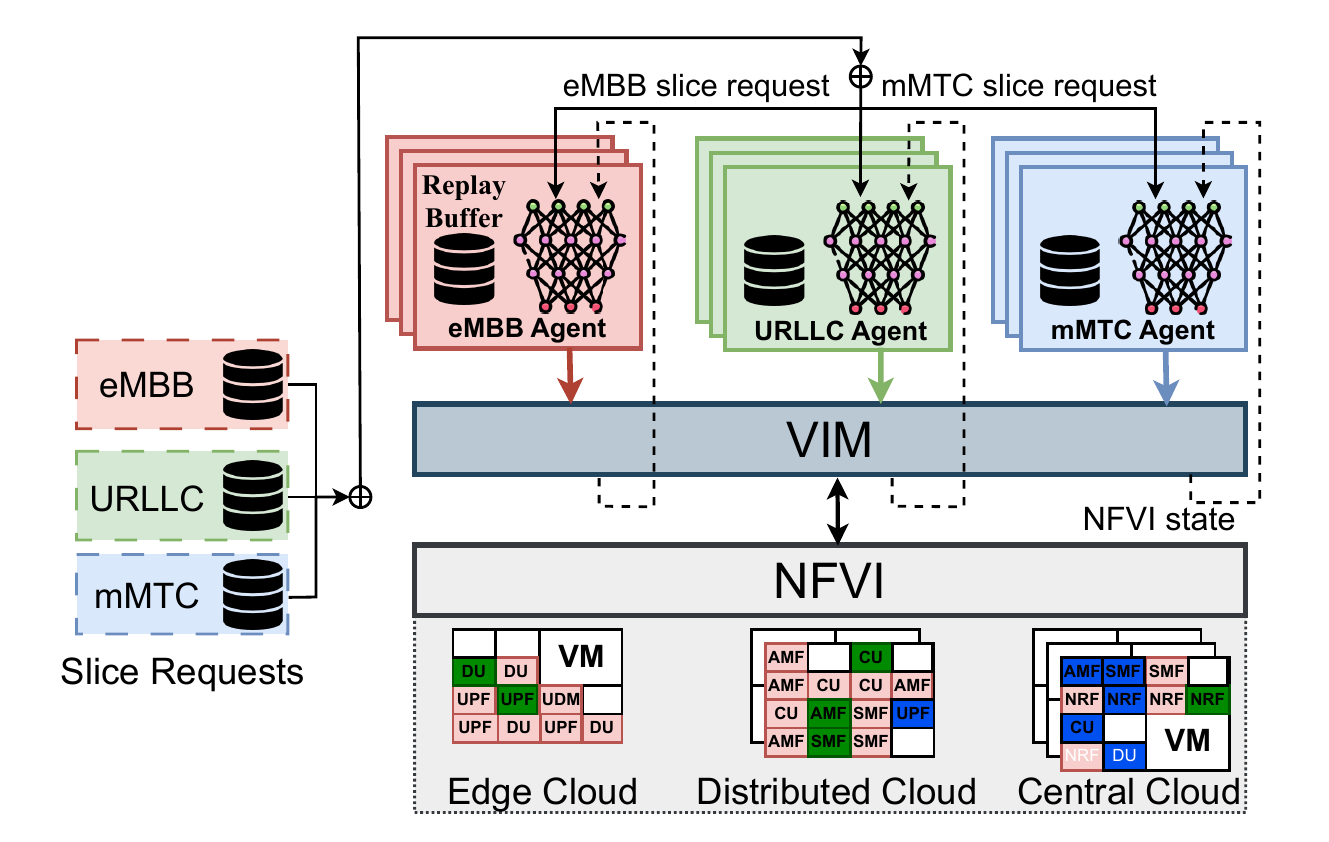} 
    \caption{High-level overview of the SPB block.}
    \label{fig:rl_scheduler}
    \vspace{-10pt}
\end{figure}

The final block of the \textit{SlicePilot} framework is a fully customized MARL scheduler, designed to optimize slice deployment within a heterogeneous multi-cloud infrastructure. It employs a disaggregated agent design to accelerate convergence, improve stability, and simplify decision-making by assigning each agent to manage slice requests of the same type. This specialization reduces the state-action space each agent needs to explore, making slice placement more efficient and scalable compared to monolithic single-agent approaches~\cite{rl6,rl7,rl8}. Operating within the NFVO, the scheduler computes placement decisions and relays them to the VIM for slice deployment and instantiation within the NFVI. The overall architecture of the SPB is illustrated in Figure~\ref{fig:rl_scheduler}.

\noindent \textbf{Optimization Framework}. We represent each network slice as an undirected graph \( G = (V,L) \), where \( V \) denotes the set of VNFs composing the slice, and \( L \) represents the set of links interconnecting them. To distinguish between functional roles, the set \( V \) is further decomposed into two subsets: \( V_u \), which contains the user plane VNFs, and \( V_c \), which contains the control plane VNFs, such that \( V = V_u \cup V_c \) and \( V_u \cap V_c = \emptyset \).  Each VNF \( v \in V \) has resource demands in terms of computation and storage, denoted by \( r_v = (r_v^{\text{cpu}}, r_v^{\text{mem}}) \). The total resource capacity of the NFVI is distributed across \( M \) different cloud infrastructures, where each infrastructure $m \in M$ has a total resource capacity \( C_m = (C_m^{\text{cpu}}, C_m^{\text{mem}}) \). VNFs collocated in the same infrastructure \( m \) experience negligible link latency, while VNFs placed in different infrastructures (\( m \neq m' \)) experience a transmission and propagation latency denoted as \( D_{m,m'}^{\text{link}} \). Finally, each cloud infrastructure \( m \) incurs a cost associated with resource usage, denoted by \( \lambda_m \), charged hourly per baseline unit of computational and storage resources. 

To model the slice placement problem, we introduce a binary decision variable \( x_{v,m} \), where \( x_{v,m} = 1 \) indicates that VNF \( v \) is placed on infrastructure node \( m \), and \( x_{v,m} = 0 \) otherwise. To ensure feasibility, we introduce the following constraints:

\noindent First, each VNF must be placed in one cloud infrastructure:
\begin{equation}
\sum_{m \in M} x_{v,m} = 1, \quad \forall v \in V
\end{equation}
\noindent Second, the total computational demand of VNFs assigned to cloud infrastructure $m$ must not exceed its capacity:
\begin{equation}
\sum_{v \in V_{\text{total}}} r_{v}^{\text{cpu}} \cdot x_{v,m} \leq C_m^{\text{cpu}}, \quad \forall m \in M
\end{equation}
\noindent Third, the total storage resources requested by VNFs in a cloud infrastructure $m$ must not exceed its storage capacity:
\begin{equation}
\sum_{v \in V_{\text{total}}} r_{v}^{\text{mem}} \cdot x_{v,m} \leq C_m^{\text{mem}}, \quad \forall m \in M
\end{equation}
\noindent Fourth, depending on the slice type \( z \in Z \), the scheduler needs to activate only the relevant slice-specific constraints from a set of available constraints. While several new constraints can be introduced, we consider two general cases applicable to most slice types: one constraint for placing the VNFs to ensure the end-to-end user plane latency, denoted as \( D_z \), is satisfied —where this latency represents the total delay experienced as traffic traverses through all user plane VNFs before reaching the application servers—, and an additional constraint for consolidating VNFs within the same cloud infrastructure to minimize inter-cloud communication overhead, which is particularly beneficial for slices that generate high volumes of signaling traffic.  The user plane latency and VNF consolidation constraints are detailed in Equations (4) and (5), respectively. We focus on these two constraints as they reflect the current enforcement capabilities of existing 5G open-source software stacks such as OAI, where more complex or fine-grained placement policies cannot yet be fully realized.
\begin{equation}
\sum_{\substack{v, v+1 \in V_z}}  
\sum_{m \in M} \sum_{m' \in M}  
D_{m,m'}^{\text{link}} \cdot x_{v,m} \cdot x_{v+1,m'} 
\leq D_z, \quad 
\end{equation}
\begin{equation}
x_{v,m} = x_{v',m}, \quad \forall v, v' \in V_z, \quad \forall z \in Z, \quad \forall m \in M
\end{equation}
\noindent Based on this formulation, the objective of the scheduler is to optimally place the VNFs of each slice to minimize the total costs associated with reserving computational and storage resources in the heterogeneous multi-cloud infrastructure:
\begin{equation}
\begin{aligned}
\min_{x} \quad & \sum_{v \in V} \sum_{m \in M} \lambda_m \cdot r_v^{\text{cpu}} \cdot r_v^{\text{mem}} \cdot x_{v,m}  \\
\text{s.t.} \quad & \text{constraints: (1) - (5)} \\
& x_{v,m} \in \{0,1\}, \quad \forall v \in V, \quad \forall m \in M
\end{aligned}
\end{equation}

\noindent The optimization problem formulated in (6) is an integer linear programming problem and a variant of the VNE problem (NP-hard). Given this complexity, obtaining an optimal solution using conventional optimization methods is computationally intractable. Therefore, to approximate the solution we adopt a learning-based heuristic approach to efficiently derive near-optimal solutions while reducing computational overhead.

\noindent \textbf{Slice-Wise Problem Decomposition}. We decompose the optimization problem described in (6) into a set of smaller optimization problems, each corresponding to a specific slice type $z \in Z$. These sub-problems can be solved independently based on slice request arrival times, i.e., the first slice request is processed first, followed by subsequent requests in their respective order of arrival. We model each sub-problem as a Markov Decision Process (MDP) \cite{mdp}, viewing it as a sequential decision-making process where VNFs are placed one at a time until all required VNFs are deployed.
Each MDP state \( s_t \in S \) is described as follows:
\begin{equation} 
s_t = \left[ \mathbf{\hat{C}}(t), Q_z(t), \mathbf{T}_z(t), \mathbf{L}_z(t) \right]
\end{equation}
\noindent \noindent where \( \mathbf{\hat{C}}(t) = (\hat{C}_1(t), \hat{C}_2(t), \dots, \hat{C}_m(t)) \in \mathbb{R}^{1 \times 2m} \) is a row vector representing the currently available computational and storage resources in each cloud infrastructure \( m \), \( Q_z(t) \in \mathbb{R} \) is a scalar denoting the total number of queued slices of type \( z \) at time step \( t \), \( \mathbf{T}_z(t) = (\sum_{q \in Q_z} \sum_{v \in V} r_v^{\text{cpu}}, \sum_{q \in Q_z} \sum_{v \in V} r_v^{\text{mem}})  \) \( \in \mathbb{R}^{1 \times 2}\) is a row vector describing the total aggregated resource demand of all queued slices of type \( z \) at time step \( t \), and \( \mathbf{L}_z(t) = (L_1(t), L_2(t), \dots, L_{Q_z}(t)) \in \mathbb{R}^{1 \times Q_z} \) is a row vector containing the SLA-related parameters for all slices of type \( z \).  The action selection process follows a common structure across all MDPs, where the action space \( a_t\in A \) is represented as a one-hot vector:
\begin{equation}
a_t = [a_1(t), a_2(t), \dots, a_m(t)] \in \{0,1\}^m
\end{equation}
\noindent where \( a_m(t) \in \{0, 1\} \) denotes the binary placement decision for VNF \( v \) on cloud infrastructure \( m \) at time step \( t \). Lastly, each MDP has its own reward function \( R: S \times A \to \mathbb{R} \), to evaluate the effectiveness of the chosen actions in achieving the optimization objective while satisfying the given constraints. Three types of rewards are assigned in each MDP:

\noindent (1) A reward \( r_1 \in \mathbb{R} \) is assigned during the placement of each VNF to penalize resource constraint violations:  
\begin{equation}
r_1(s_t, a_t) =
\begin{cases}
    \delta_1, & \text{if } a_t \text{ violates resource constraints}  \\
    0, & \text{otherwise}.
\end{cases}
\end{equation}
\noindent where \( \delta_1 \) is a negative scalar (\(\delta_1 < 0\)).

\noindent (2) A reward \( r_2 \in \mathbb{R} \) is assigned during the placement of each VNF to encourage cost-efficient decisions:
\begin{equation}
r_2(s_t, a_t) = \boldsymbol{\delta_2} \cdot \mathbf{I}(a_t)
\end{equation}
where \( \boldsymbol{\delta_2} = [\delta_2(1), \delta_2(2), \dots, \delta_2(M)]\) is a predefined row vector that assigns a cost-efficiency reward \( \delta_2(m) \) to each selected infrastructure \( m \), and \( \mathbf{I}(a_t) \) is a one-hot column vector indicating the selected infrastructure $m$ at time step $t$. 

\noindent (3) A reward \( r_3 \in \mathbb{R} \) is given at the end of the slice placement process for each slice to ensure compliance with SLAs:  
\begin{equation}
r_3(s_t, a_t) =
\begin{cases}
    \delta_3, & \text{if the slice's SLAs are met}, \\
    0, & \text{otherwise}.
\end{cases}
\end{equation}
\noindent where \( \delta_3 \) is a positive scalar (\(\delta_3 > 0\)). Therefore, the total reward \( R_t(s_t, a_t) \) at each time step is the sum of the individual rewards assigned during the slice placement process:
\begin{equation}
R_t(s_t, a_t) = r_1(s_t, a_t) + r_2(s_t, a_t) + r_3(s_t, a_t).
\end{equation}

\noindent \textbf{Learning Slice-Wise Placement Policies}. To solve each formulated MDP, we employ RL techniques to determine the optimal sequence of actions that maximize the expected cumulative rewards.  Specifically, for each MDP, the objective is to learn a dedicated policy—mapping states to actions—that maximizes the accumulated time-discounted rewards, defined as \( \pi^* = \arg\max_{\pi} \mathbb{E} \left[ \sum_{t=0}^{\infty} \gamma^t R_t(s_t,a_t) \right] \), where \( \gamma \in (0,1] \) is the discount factor regulating the importance of future rewards. Each MDP is handled by a corresponding agent, which interacts with its environment by observing states, selecting actions, and receiving feedback in the form of rewards. Through this continuous interaction, each agent gradually refines its strategy, ultimately converging to an optimal policy tailored to its slice-specific placement problem.

A key challenge in this formulation is how agents retain and utilize past experiences. Classical RL methods often rely on tabular approaches to store and update experiences, but these become impractical in large state-action spaces due to limited generalization~\cite{sutton2018reinforcement}. In our case, the state-action space per agent grows exponentially as \( \mathcal{O}(|V| \cdot |M| \cdot |Q_z| \cdot |M|^{|V|}) \), necessitating scalable learning techniques for efficient generalization. Function approximation has been explored to address this issue~\cite{bertsekas}, with Deep Neural Networks (DNNs) emerging as a promising solution due to their success in recent applications~\cite{rl7, rl8}. Motivated by this, we employ DNNs as value function approximators within our agents to cope with the high-dimensional state-action space.

\noindent \textbf{Deep Q-Learning for Value Function Approximation.}  
To enable efficient decision-making in each agent, we adopt \textit{Deep Q-Learning} (DQL), an RL approach that integrates Q-learning with DNNs to generalize over high-dimensional state-action spaces~\cite{qlearning, dqn}. The goal is to learn a Q-function that maps each state-action pair \( (s, a) \) to the expected cumulative reward obtained by taking action \( a \) in state \( s \) and following a policy thereafter, formally defined as \( Q: \mathcal{S} \times \mathcal{A} \to \mathbb{R} \).

The optimal Q-function, denoted as \( Q^* \), satisfies the Bellman equation, which recursively relates the value of a state-action pair to the immediate reward and the discounted value of the next state:
\begin{equation}
Q^*(s, a) = \mathbb{E}_{s'} \left[ r(s, a, s') + \gamma \max_{a'} Q^*(s', a') \right],
\end{equation}
\noindent where \( s' \) is the next state, \( a' \) is the next action, and the expectation is taken over the transition distribution \( P(s'|s,a) \). To approximate \( Q^*(s,a) \), we use a DNN parameterized by \( \theta \), denoted as \( Q_{\theta}(s,a) \). The agent iteratively updates this network through a value iteration process that refines its reward estimates over time~\cite{qlearning}. This is achieved by minimizing the difference between the predicted and target Q-values using a mean squared error loss:
\begin{equation}
L(\theta) = \mathbb{E} \left[ (y - Q_{\theta}(s,a))^2 \right],
\end{equation}
\noindent where the target value \( y \) is computed as:
\begin{equation}
y = r(s,a,s') + \gamma \max_{a'} Q_{\theta}(s', a').
\end{equation}

\noindent To optimize \( Q_{\theta}(s,a) \), the parameters \( \theta \) are updated via gradient descent:
\begin{equation}
\theta \leftarrow \theta - \alpha \nabla_{\theta} L(\theta),
\end{equation}
\noindent where \( \alpha \) is the learning rate and \( \nabla_{\theta} L(\theta) \) denotes the gradient of the loss with respect to the network parameters. Unlike supervised learning, the target in DQL depends on the same network being trained, which can lead to instability during learning. To address this, we apply the following enhancements to improve training stability and sample efficiency:

\begin{enumerate}
    \item \textbf{Target Network.} We maintain a delayed version of the Q-function, \( Q_{\theta'}(s,a) \)~\cite{dqn}, to compute the target value more reliably. The target in (15) is modified as:
    \begin{equation}
    y = r(s, a, s') + \gamma \max_{a'} Q_{\theta'}(s', a'),
    \end{equation}
    and the target network is updated every \( C \) steps by copying weights from the current network.

    \item \textbf{Double Q-Learning.} To mitigate the overestimation bias caused by using the same Q-function for both action selection and evaluation, we use Double Q-learning~\cite{ddqn}. The action that maximizes the Q-value is selected using the current network:
    \begin{equation}
    a^* = \arg\max_{a'} Q_{\theta}(s', a'),
    \end{equation}
    while the corresponding Q-value is evaluated using the target network:
    \begin{equation}
    y = r(s, a, s') + \gamma Q_{\theta'}(s', a^*).
    \end{equation}

    \item \textbf{Experience Replay.} To improve sample efficiency and reduce temporal correlations, each agent maintains an experience replay buffer \( \mathcal{I} \) that stores past transitions \( (s, a, r, s') \). Instead of updating the network using only the most recent experience, we sample a mini-batch of \( B \) transitions from \( \mathcal{I} \) during each training iteration. The buffer holds up to \( N \) transitions and enables more stable and data-efficient learning.
\end{enumerate}

\section{EXPERIMENTATION ENVIRONMENT} \label{sec:experimentation}
In this section, we present the experimental setup used to realize the functional blocks of \textit{SlicePilot}.

\begin{figure}[t]
    \centering
    \begin{minipage}{0.45\linewidth}
        \centering
        \includegraphics[width=0.95\linewidth]{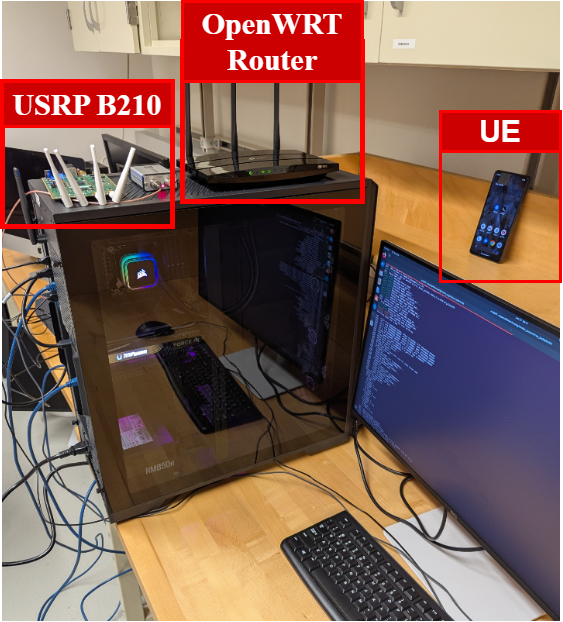}
    \end{minipage}
    \hspace{0.05\linewidth} 
    \begin{minipage}{0.45\linewidth}
        \centering
        \includegraphics[width=0.95\linewidth]{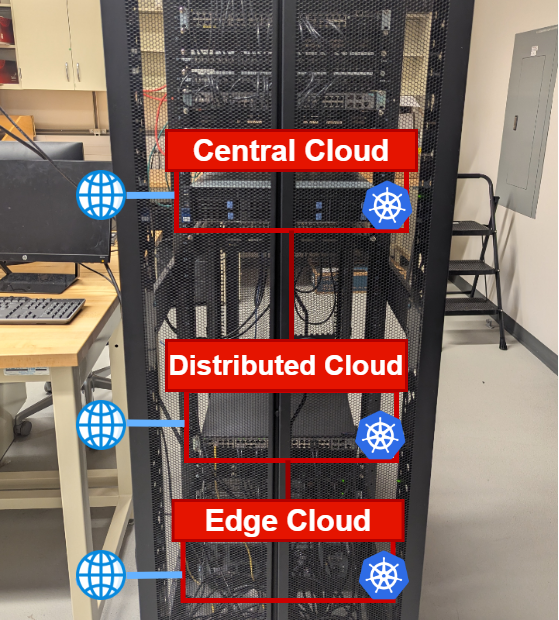}
    \end{minipage}

    \vspace{-3pt}
    \caption{Experimental setup: (left) Over-the-air 5G testbed, (right) heterogeneous multi-cloud infrastructure.}
    \label{fig:testbed_overview}
    \vspace{-12pt}
\end{figure}

\noindent \textbf{User Plane Traffic Collection.} For the traffic collection campaign, we leveraged our in-lab over-the-air 5G standalone testbed, which operated in the n78 TDD band with a 40 MHz bandwidth and utilized a USRP B210 SDR as the RF front end, as illustrated in Figure~\ref{fig:testbed_overview}. The baseband processing and core network VNFs were deployed as Docker containers using the latest OAI release \cite{oai}. A Google Pixel 7 mobile device served as the primary UE for most network scenarios. Additionally, to connect external devices, we tethered the Pixel to an OpenWRT router acting as a gateway to our 5G network. We considered two external devices: an AR/VR headset (Microsoft HoloLens 2) and a Govee WiFi Thermometer Hygrometer H5179 smart humidity and temperature IoT sensor.

Building on this setup, we conducted a series of traffic collection experiments targeting three slice types: URLLC, eMBB, and mMTC. For each application scenario, we launched the base station (gNB), verified its connection to the core network, and connected the UE by disabling airplane mode. Once the SMF assigned an IP address, we captured traffic on the gNB host using \textit{tcpdump} \cite{tcpdump}, filtering by the UE’s IP to isolate user-plane packets and exclude control signaling. Each scenario was recorded for 4 hours to ensure sufficient traffic volume and diversity. The application use cases were grouped by slice type: URLLC included VoIP (Messenger, WhatsApp), online gaming (Fortnite, PUBG), web browsing, and AR/VR (Microsoft Dynamics 365 Remote Assist); eMBB included media streaming and downloads (YouTube, TikTok), social media (Facebook, Instagram), and video conferencing (Messenger, Skype, Zoom); mMTC focused on IoT traffic.

\noindent \textbf{Large-Scale VNF Stress Testing.} To enable large-scale slice instantiation and traffic replay, we replaced the physical USRP and devices with OAI’s emulated RF front ends and NR-UEs. For each network slice, we instantiated dedicated user plane VNFs and configured static routes between them to ensure correct forwarding across the data path. We also customized the NR-UE and traffic server containers to replay only the relevant traffic profiles corresponding to each slice type. To support concurrent high-volume traffic flows, we implemented a lightweight UDP server within each VNF, binding it to multiple ports to handle parallel connections from emulated devices. During each experiment, we replayed the collected traffic traces through the instantiated slices, ensuring that each slice carried its own application-specific load. Throughout the experiments, we monitored VNF-level resource consumption metrics—including CPU and RAM usage—under varying connection loads to analyze the relationship between traffic intensity and compute demand. All experiments were conducted on a commercial-grade server equipped with an AMD EPYC 7352 24-core processor and 128 GB of DDR4 RAM.

\noindent \textbf{Multi-Cloud Infrastructure and Scheduler Deployment.} To evaluate our scheduler in a real deployment, we set up a multi-cloud infrastructure with OpenStack as the VIM to manage computing, storage, and networking resources. A Kubernetes cluster was deployed on top of OpenStack to serve as a simplified NFVO, orchestrating network slices and handling their lifecycle. The MARL scheduler was integrated within the NFVO, interacting with the environment by receiving state updates from the Kubernetes Metrics Server and issuing placement decisions via the Kubernetes API Server. As illustrated in Figure~\ref{fig:testbed_overview}, the infrastructure was deployed across three COTS servers, each equipped with an AMD EPYC 7352 24-core processor, 128 GB DDR4 RAM, and four NVIDIA RTX A5000 GPUs. These servers served as computing nodes representing the edge, distributed, and central cloud infrastructures. Since all servers were physically colocated within the same rack, kernel modifications were applied to introduce latencies, accurately emulating the latency characteristics of a geographically distributed infrastructure.


\section{EVALUATION RESULTS} \label{sec:evaluation}

\subsection{Configurations and Settings}
\noindent \textbf{Infrastructure Configurations}. We consider \( M = 3 \) heterogeneous cloud infrastructures—edge, distributed, and central—and set their resource capacities as follows: \( C_0= \) (16 CPU cores, 16 GiB RAM), \( C_1 = \) (32 CPU cores, 32 GiB RAM), and \( C_2 = \) (64 CPU cores, 64 GiB RAM), respectively. Each infrastructure hosts virtual machines instantiated using predefined flavors: (1 CPU core, 1~GiB RAM), (2 CPU cores, 2~GiB RAM), and (4 CPU cores, 4~GiB RAM). We select cost values to reflect the hourly charge for on-demand reservation of 1 CPU core and 1 GiB RAM, set as \( \lambda_0 = 0.010 \) \$/h at the edge, \( \lambda_1 = 0.005 \) \$/h at the distributed cloud, and \( \lambda_2 = 0.001 \) \$/h at the central cloud, following commercial pricing trends~\cite{aws_ec2_pricing}. Inter-cloud latencies are modeled as normal distributions, following values reported in~\cite{atalay2023first}: edge-to-distributed latency \( D_{0,1}^{\text{link}} \sim \mathcal{N}(0.5, 0.1^2) \, \text{ms} \), and distributed-to-central \( D_{1,2}^{\text{link}} \sim \mathcal{N}(20, 1^2) \, \text{ms} \). Finally, latency to external networks depends on UPF placement—5~ms at the edge, 7.5~ms in the distributed cloud, and 10~ms in the central cloud.

\noindent \textbf{Slice Configurations}. We represent each slice with dedicated VNFs, including NRF, UDR/UDM/AUSF (U/U/A), AMF, SMF, UPF, CU, and DU. As a proof of concept, we assume uniform resource demands across  slices using the default values provided by OAI. NRF is assigned \( r_0 = (0.15 \) CPU cores, \( 0.128 \) GiB RAM\() \),  UDR/UDM/AUSF requires \( r_1 = (0.65 \) CPU cores, \( 0.896 \) GiB RAM\() \). AMF and SMF are allocated \( r_3 = (0.25 \) CPU cores, \( 0.256 \) GiB RAM\() \) and \( r_4 = (0.25 \) CPU cores, \( 0.256 \) GiB RAM\() \), respectively. Similarly, UPF and CU are assigned \( r_5 = (0.5 \) CPU cores, \( 0.512 \) GiB RAM\() \) and \( r_6 = (0.5 \) CPU cores, \( 0.512 \) GiB RAM\() \), respectively, while the DU requires \( r_7 = (3 \) CPU cores, \( 2 \) GiB RAM\() \). Finally, we define latency constraints for each slice type. For URLLC slices, we set the end-to-end user plane latency requirement to \( D_{\text{URLLC}} < 10 \) ms, while for eMBB and mMTC slices, \( D_{\text{eMBB}} < 20 \) ms and \( D_{\text{mMTC}} < 50 \) ms, respectively.


\noindent \textbf{Training and RL Configurations}.  
All \textit{SlicePilot} agents are trained using a shared configuration, and the reported values are obtained through hyperparameter optimization: batch size \( B = 32 \), replay buffer size \( \mathcal{I} = 20{,}000 \), number of episodes \( E = 50{,}000 \), and synchronization interval \( C = 1000 \). An exploration decay of \( \lambda_{\epsilon} = 25{,}000 \) is used for all agents. Each agent employs the ReLU activation function, with three hidden layers of size 128 for eMBB and URLLC, and 256 for mMTC. The learning rate is set to 0.05 for eMBB, 0.01 for URLLC, and 0.005 for mMTC, while the discount factor \( \gamma \) remains 0.01 across all agents. The reward function parameters are set as follows: \( \delta_1 = -100 \), \( \delta_2 = [1, 2, 4] \), and \( \delta_3 = 15 \), 20, and 10 for eMBB, URLLC, and mMTC, respectively.

\begin{figure*}[t]
    \centering
    \begin{subfigure}{0.31\textwidth}
        \centering
        \includegraphics[width=0.94\textwidth]{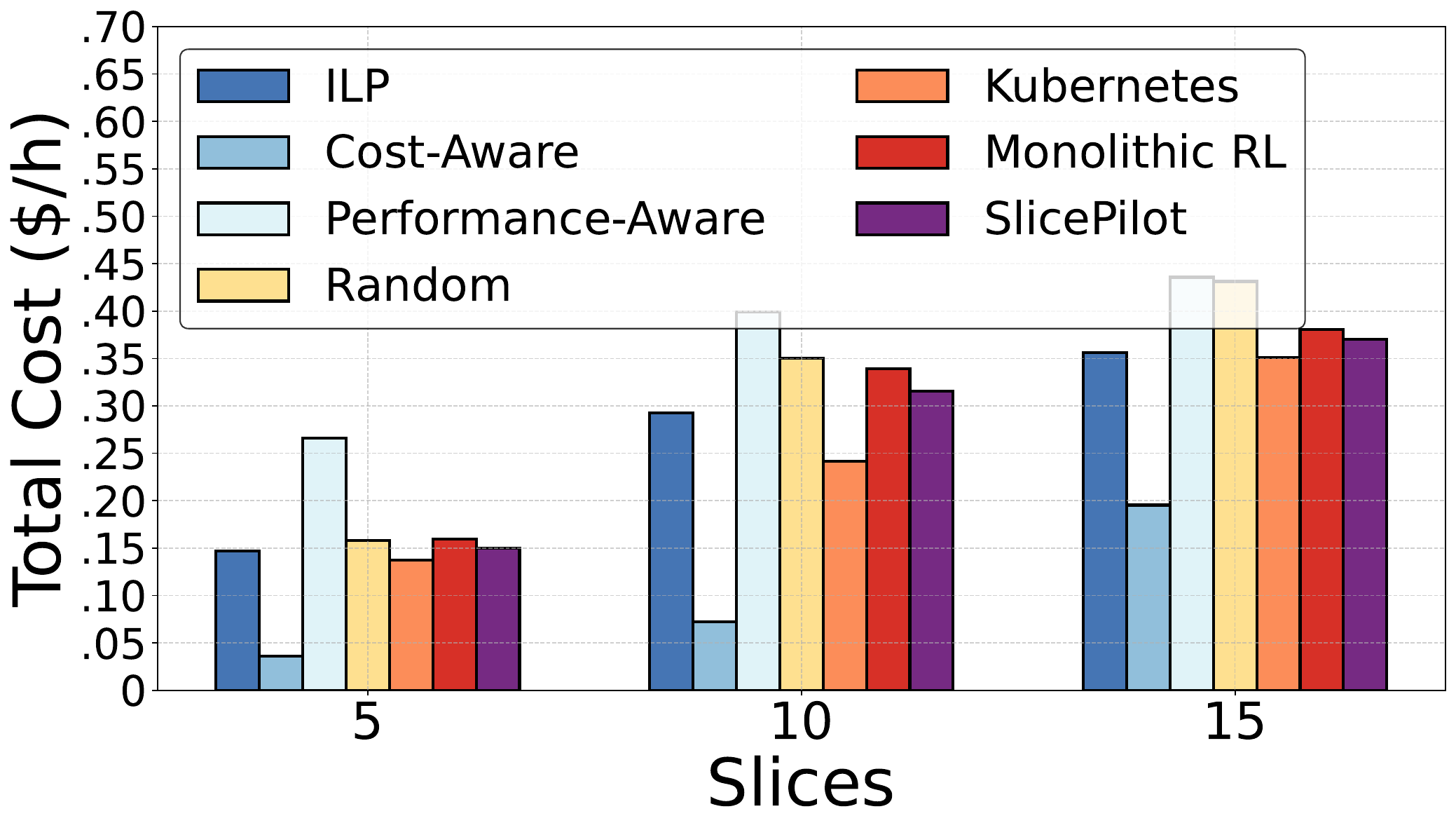}
        \vspace{-7pt} 
        \caption{Total deployment cost.}
    \end{subfigure}
    \begin{subfigure}{0.31\textwidth}
        \centering
        \includegraphics[width=0.94\textwidth]{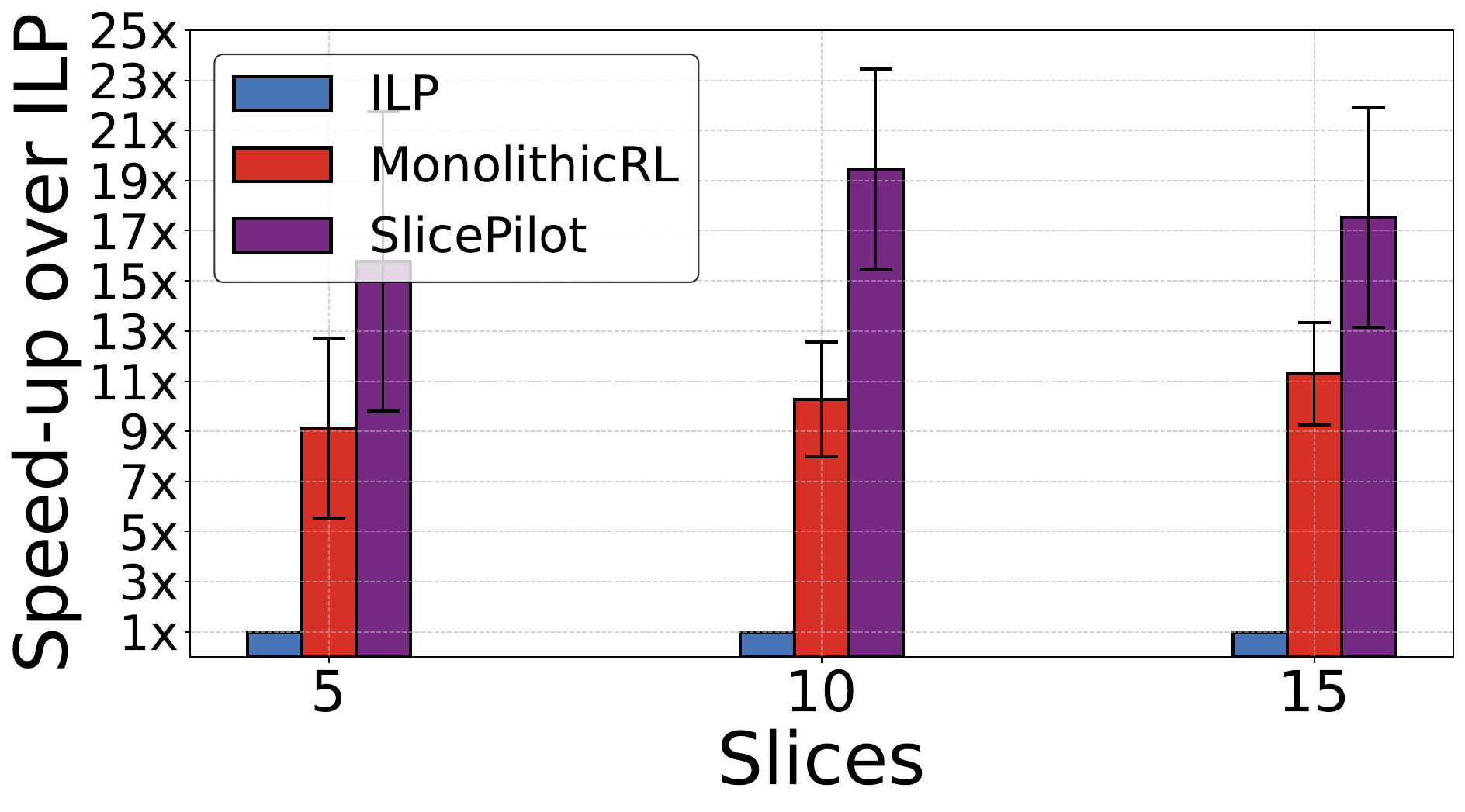}
        \vspace{-7pt} 
        \caption{Execution time speed-up.}
    \end{subfigure}
    \begin{subfigure}{0.31\textwidth}
        \centering
        \includegraphics[width=0.94\textwidth]{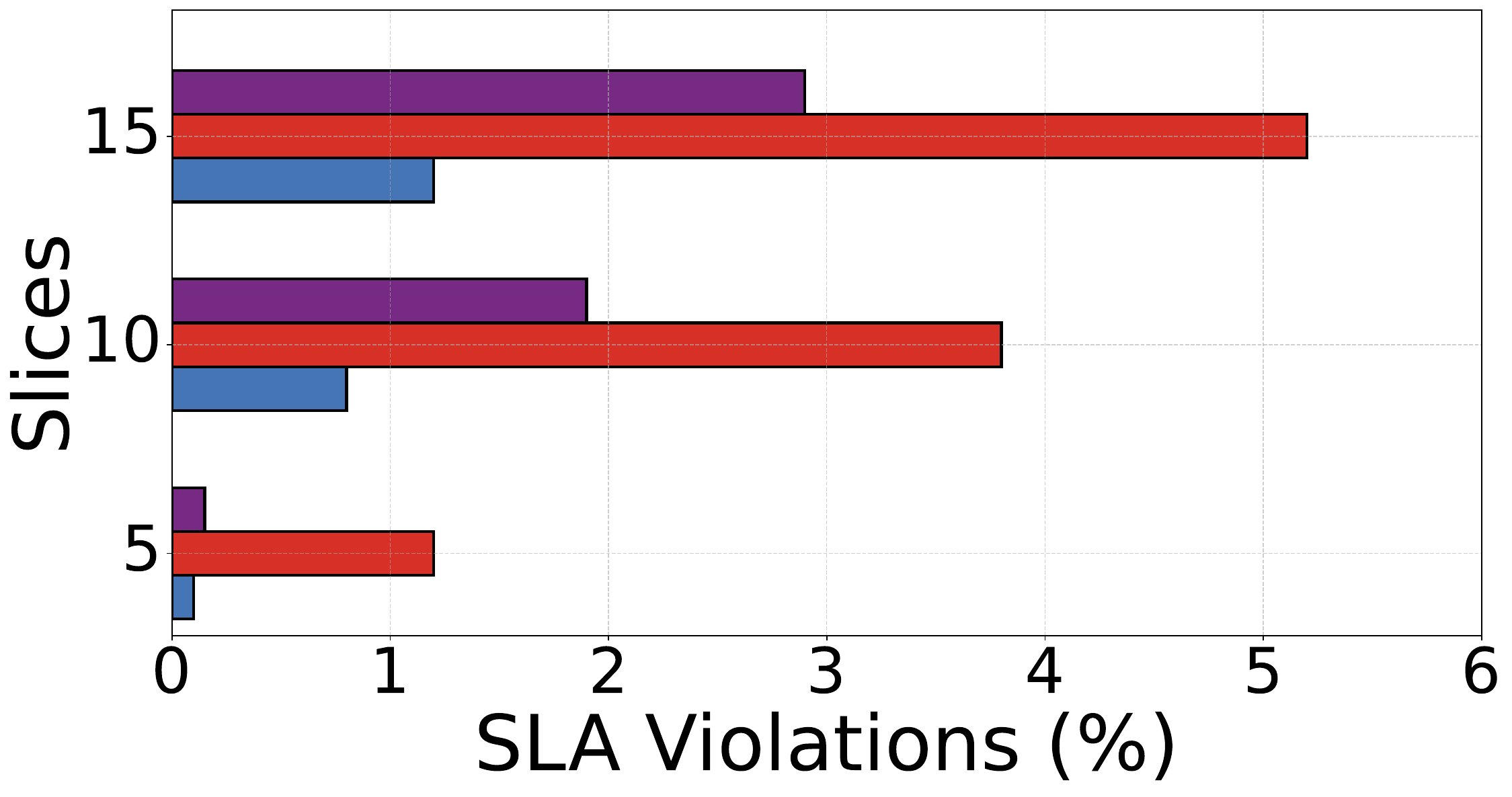}
        \vspace{-7pt} 
        \caption{SLA violations percentage.}
    \end{subfigure}

    \vspace{10pt}
    
    \begin{subfigure}{0.47\textwidth}
        \centering
        \includegraphics[width=0.94\textwidth]{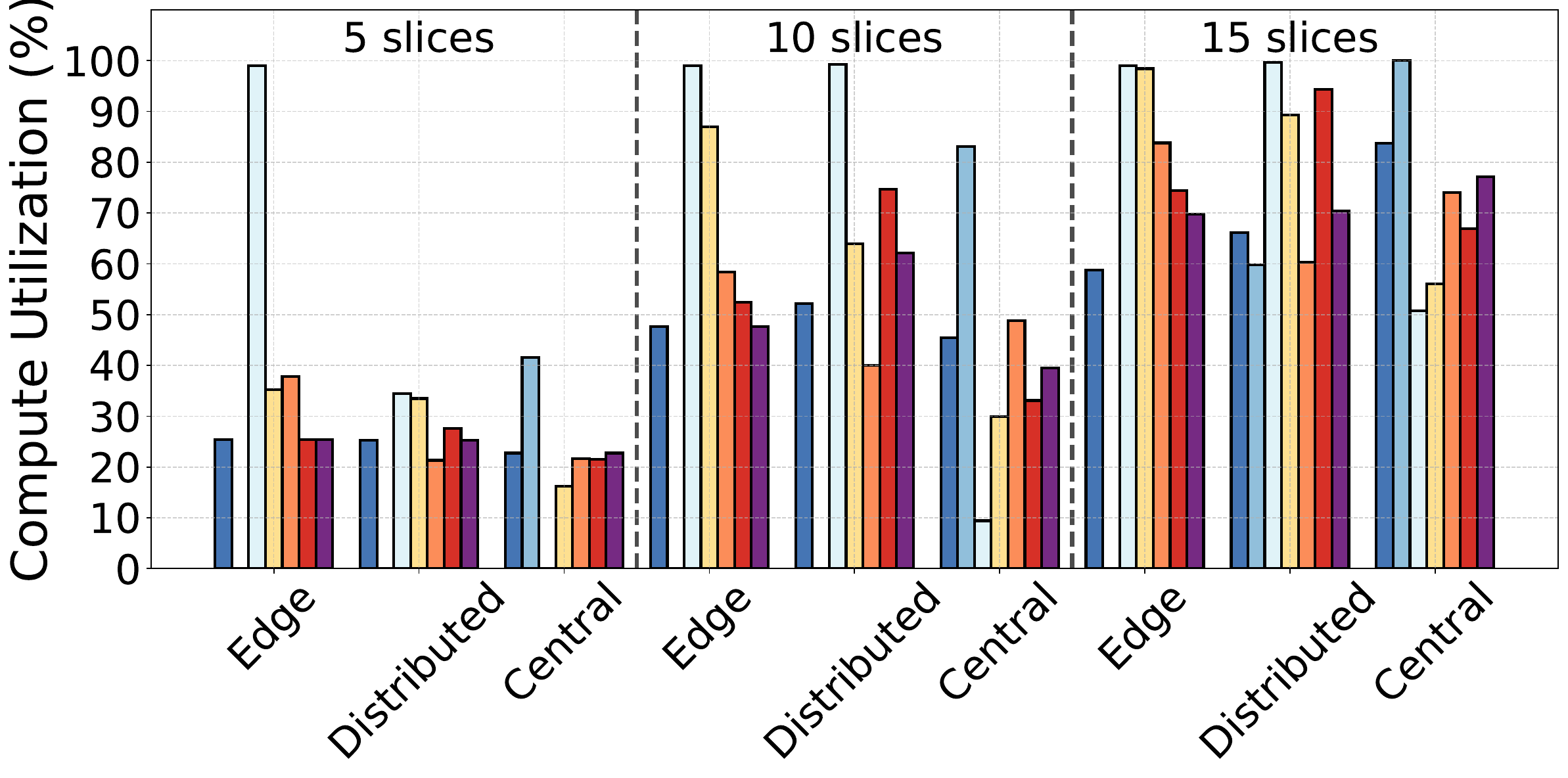}
        \vspace{-7pt}
        \caption{CPU utilization.}
    \end{subfigure}
    \begin{subfigure}{0.47\textwidth}
        \centering
        \includegraphics[width=0.94\textwidth]{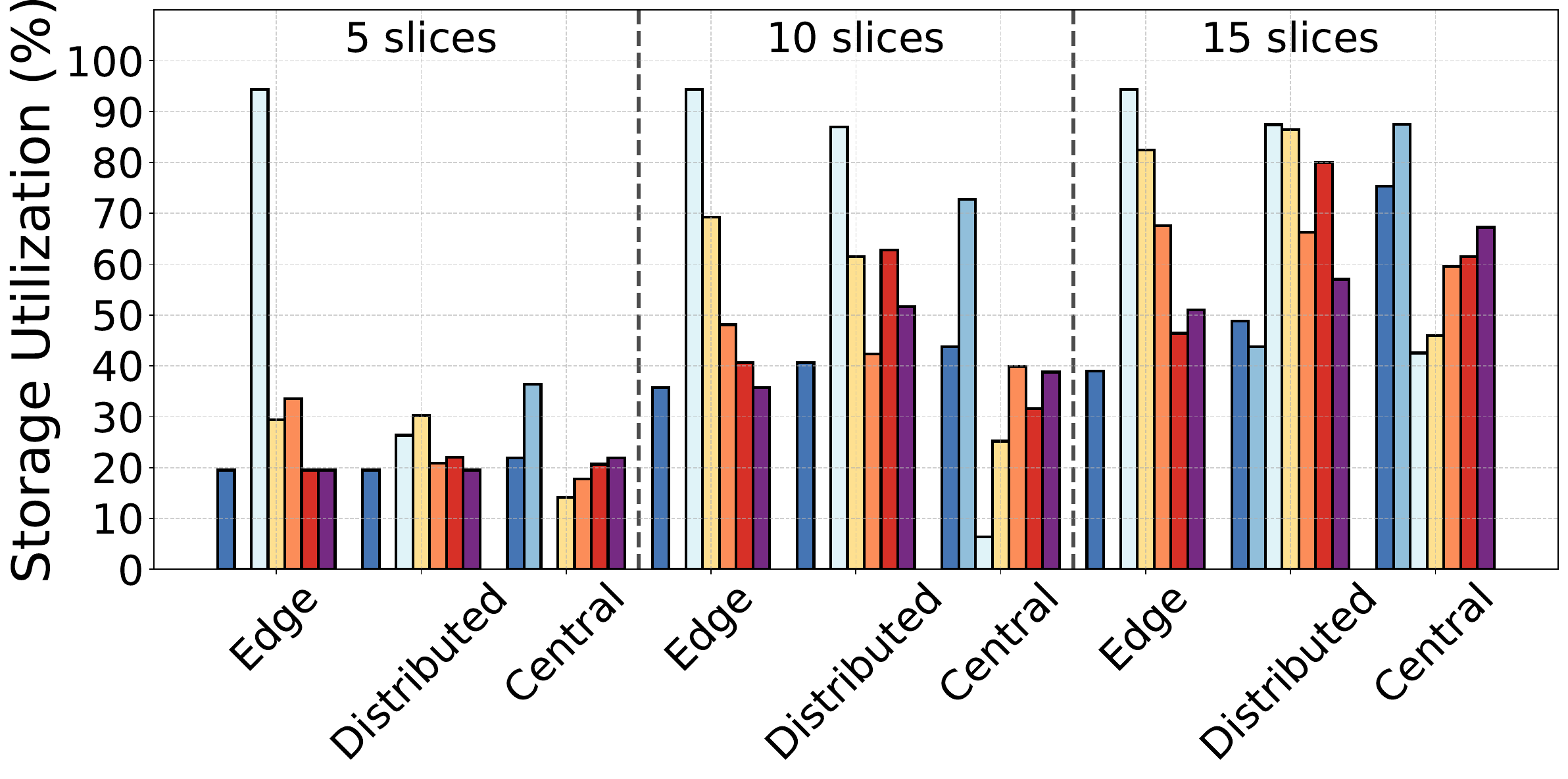}
        \vspace{-7pt} 
        \caption{Storage utilization.}
    \end{subfigure}
    \caption{Performance evaluation of \textit{SlicePilot} and baseline placement algorithms in terms of (a) deployment cost, (b) execution time speed-up, (c) SLA violation rate, (d) CPU utilization, and (e) storage utilization.}
    \label{fig:evaluation_results}
\end{figure*}

\subsection{Placement Algorithms}
We compare \textit{SlicePilot}’s disaggregated scheduler against a set of baseline placement algorithms:
 \\
\noindent \textbf{1) ILP Algorithm}. Solves the optimization problem in (6) to obtain optimal placement but is computationally expensive. We use it as an \textit{upper bound} on placement performance. A simplified version is employed by OSM MANO~\cite{osm}.
\\
\noindent \textbf{2) Cost-Aware Algorithm}. A heuristic that places slices in the lowest-cost cloud tiers until resources are exhausted.
\\
\noindent \textbf{3) Performance-Aware Algorithm}. A heuristic that places slices on infrastructures closer to users, shifting to higher-tier clouds as local resources deplete.
\\
\noindent \textbf{4) Random Algorithm}. A heuristic that assigns slice placements arbitrarily, without optimization criteria.
\\
\noindent \textbf{5) Load Balancing Algorithm (Kubernetes Scheduler)}. The default Kubernetes scheduler, which places VNFs based on resource availability to balance utilization across the multi-cloud infrastructure~\cite{k8s_scheduler}.
\\
\noindent \textbf{6) Monolithic RL Algorithm}: A single-agent RL approach for all slice placement decisions. It uses the same training setup as \textit{SlicePilot}, with a 3-layer DNN (128 neurons/layer), ReLU activation, learning rate \( \alpha = 0.05 \), and discount factor \( \gamma = 0.01 \), selected via hyperparameter optimization.

\noindent While Algorithms 2)–5) enable fast placement decisions, they ignore slice-specific SLAs. Thus, we evaluate Algorithms 1), 6), and \textit{SlicePilot} based on execution time and SLA compliance, and use Algorithms 2)–5) to assess resource utilization and deployment cost.

\subsection{Experimental Results}
We evaluate \textit{SlicePilot} based on computational efficiency, cost, resource utilization, and SLA compliance. To test adaptability under varying slice requests, we consider deployments of 5, 10, and 15 slices, corresponding to 25\%, 50\%, and 75\% of total infrastructure capacity. Slice requests—URLLC, eMBB, and mMTC—are sampled per scenario from a multinomial distribution with probabilities 0.2, 0.3, and 0.5, respectively, to ensure diversity in slice types. Each scenario is executed multiple times, and the reported results reflect the average performance across all trials.

\noindent \textbf{Cost Efficiency.} We evaluate the hourly deployment cost (\$/h) of network slices across placement strategies. As shown in Figure~\ref{fig:evaluation_results}(a), \textit{SlicePilot} consistently achieves near-optimal cost, closely matching the ILP baseline. At 25\%, 50\%, and 75\% load, \textit{SlicePilot}'s costs are only 2.04\%, 7.8\%, and 4.1\% higher than ILP, respectively. In contrast, the Monolithic RL incurs notably higher costs due to centralized decision-making, with up to 15.8\% overhead at medium load.  Heuristic strategies exhibit a trade-off: the Cost-Aware algorithm minimizes cost but suffers from high SLA violations, while Performance-Aware prioritizes SLAs at a significantly higher cost—up to 81\% above ILP. Load Balancing offers moderate savings but lacks SLA-awareness, and Random placement performs the worst overall. Notably, deployment costs do not scale proportionally with the number of slices. This is primarily due to the heterogeneous nature of cloud resources, and the pricing differences across cloud tiers.

\noindent \textbf{Computational Efficiency.} Second, we evaluate the computational efficiency of different slice placement strategies relative to ILP, defined as the time required to compute placement decisions upon slice request arrival. Figure~\ref{fig:evaluation_results}(b) presents the speed-up achieved by Monolithic RL and \textit{SlicePilot} compared to ILP for different number of slice requests. As expected, ILP exhibits the highest execution time due to its exhaustive search process, making it computationally prohibitive for large-scale slice deployments. In contrast, Monolithic RL achieves a speed-up of approximately 9x for 5 slices, increasing to 11x for 10 slices and 10x for 15 slices. \textit{SlicePilot} further improves computational efficiency, demonstrating a speed-up of approximately 16x for 5 slices, 19x for 10 slices, and 17x for 15 slices. This reduction in execution time is due to the agents' learned ability to apply optimal placement decisions, thereby avoiding brute-force combinatorial searches during inference. Overall, \textit{SlicePilot} achieves up to 73\% faster execution than Monolithic RL and maintains an order-of-magnitude improvement over ILP, making it a practical and scalable solution for real-time slice placement. However, \textit{SlicePilot} does not achieve a strict $\times 3$ improvement over Monolithic RL due to the uneven distribution of slices among its schedulers, resulting from the multinomial sampling, coupled with variations in the DNN architecture of each agent.

\noindent \textbf{SLA Compliance.}  Third, we assess how different slice placement strategies uphold compliance with predefined SLAs. As shown in Figure~\ref{fig:evaluation_results}(c), SLA violation levels vary significantly across strategies. ILP achieves the highest level of compliance, with SLA violations remaining at only 0.1\%, 0.8\%, and 1.2\% for deployments of 5, 10, and 15 slices, respectively. This is expected, as ILP enforces SLA constraints during slice placement decisions. \textit{SlicePilot} demonstrates a strong balance between SLA compliance and deployment efficiency, achieving significantly fewer violations than Monolithic RL while closely approximating ILP. Across deployments, \textit{SlicePilot} maintains violations at 0.15\%, 1.9\%, and 2.9\% for 5, 10, and 15 slices, respectively. Compared to ILP, \textit{SlicePilot} incurs 1.5x times more violations for small deployments, 2.375x times more for medium deployments, and 2.42x times more for large deployments. Nevertheless, \textit{SlicePilot} achieves substantially better SLA compliance than Monolithic RL. In contrast, Monolithic RL exhibits a significantly higher percentage of SLA violations due to its centralized decision-making process. SLA violations reach 1.2\%, 3.8\%, and 5.2\% for deployments of 5, 10, and 15 slices, respectively. Finally, while Load Balancing (Kubernetes) offers lower cost, as illustrated in Figure~\ref{fig:evaluation_results}(a), its SLA compliance remains capped at 80\%, making it less viable than RL-based strategies and, for that reason, it is not included in Figure~\ref{fig:evaluation_results}(c).

\noindent \textbf{Multi-Cloud Resource Footprint and Utilization.} Finally, we analyze how different slice placement strategies allocate compute and storage resources across the edge, distributed, and central cloud infrastructures as the number of slices increases. As illustrated in Figures~\ref{fig:evaluation_results}(d),(e) resource utilization patterns differ significantly based on the placement strategy, particularly in how VNFs are distributed across cloud tiers.

\textit{SlicePilot} demonstrates balanced resource utilization across all cloud layers. For 5 slices, it distributes 25.4\% of CPU load and 19.5\% of RAM usage across the edge and distributed clouds while utilizing 22.7\% CPU and 21.9\% RAM in the central cloud. As demand increases, \textit{SlicePilot} efficiently scales resource allocation, reaching 47.6\% CPU and 35.7\% RAM utilization across edge and distributed tiers for 10 slices, while maintaining 39.5\% CPU and 38.8\% RAM in the central cloud. At 15 slices, \textit{SlicePilot} sustains 69.8\% CPU and 51.0\% RAM in edge and distributed clouds, offloading from the central cloud, which reaches 77.2\% CPU and 67.2\% RAM. These results indicate that \textit{SlicePilot} effectively balances workload distribution, reducing over-reliance on expensive edge cloud resources. In contrast, ILP achieves the lowest cost but heavily loads the central cloud, reaching 83.8\% CPU and 75.3\% RAM utilization for 15 slices, reflecting its tendency to exhaust low-cost resources first. Compared to \textit{SlicePilot}, Monolithic RL exhibits greater centralization, with CPU usage in the central cloud rising from 21.6\% (5 slices) to 66.9\% (15 slices), and RAM reaching 61.5\%. Among heuristic strategies, Cost-Aware minimizes edge and distributed usage, saturating the central cloud CPU at 99.9\% for 15 slices. Performance-Aware favors edge placement, pushing edge CPU to 99.0\% while underutilizing the central cloud. Load Balancing (Kubernetes) distributes load more evenly, with central cloud usage at 74.4\% CPU and 61.5\% RAM. The Random strategy results in unpredictable slice placement, leading to 56.1\% CPU and 45.9\% RAM utilization in the central cloud.

\section{RELATED WORK}
The problem of joint resource allocation and slice placement has been extensively studied and is often modeled as a variant of the virtual network embedding problem~\cite{surv6}. Several surveys address slice lifecycle management and placement optimization~\cite{surv1,surv3}, covering techniques for resource allocation, performance guarantees, and network efficiency. Optimization-based methods provide optimal decisions under latency, isolation, and resource constraints~\cite{opt1,opt2}, but are computationally prohibitive for large-scale deployments. Heuristic approaches reduce complexity~\cite{opt3,opt4} but struggle with dynamic network conditions and lack real-world validation. More recently, reinforcement learning has been used to improve placement scalability and adaptability~\cite{rl1,rl2,rl3,rl4}, though most studies remain limited to simulation settings, hindering their applicability in real 5G systems.

Despite 5G-compliant testbeds supporting large-scale slice deployment~\cite{testbed1,testbed2,testbed3}, an optimization engine for intelligent slice placement remains absent. Existing efforts focus primarily on scaling slices rather than optimizing their placement~\cite{scal2,scal3}. For example,~\cite{scal2} introduces a slice-as-a-service framework and evaluates deployments on local testbeds with cloud cost estimations, but without optimizing placement. Similarly,~\cite{scal3} explores slice scaling across various topologies and use cases, yet lacks mechanisms for intelligent placement. While these studies validate slice deployment and scaling, they overlook optimization under real-world constraints. To our knowledge, no existing framework jointly addresses slice placement, traffic profiling, resource demand estimation, and RL-based optimization in a real heterogeneous multi-cloud 5G setup—highlighting a key gap in the literature.

\section{CONCLUSION}
This paper presents \textit{SlicePilot}, a comprehensive framework for intelligent slice placement in heterogeneous multi-cloud infrastructures. Built on three functional blocks, \textit{SlicePilot} seamlessly integrates (i) the User Plane Traffic Capturing Block to profile and replicate real-world 5G traffic patterns, (ii) the User Plane VNF Stressing Block to accurately estimate resource demands across slices, and (iii) the Slice Placement Block, a disaggregated multi-agent RL scheduler that automates VNF deployment while minimizing costs and ensuring SLA compliance. Through extensive experimentation on a self-managed multi-cloud testbed, \textit{SlicePilot} delivers near-optimal placement efficiency, achieving deployment costs within 7.8\% of ILP, 19× faster execution, and comparable SLA compliance, with only 2.42× more violations under high-load conditions. Unlike heuristic approaches, \textit{SlicePilot} intelligently balances workload distribution, preventing costly edge overutilization while avoiding central cloud congestion, ensuring efficient, scalable, cost-effective deployments. 

\clearpage
\bibliographystyle{IEEEtran}
\bibliography{references}

\end{document}